\documentclass[12pt,preprint]{aastex}

\shorttitle{\ion{O}{1}, \ion{Ca}{2}, and \ion{Fe}{2} Emission Region in AGNs}
\shortauthors{Matsuoka et al.}

\begin{document}


%\title{Exploring \ion{Fe}{2} Emission Region by \ion{O}{1} and \ion{Ca}{2} Lines\\ in Active Galactic Nuclei}
\title{Observations of \ion{O}{1} and \ion{Ca}{2} Emission Lines in Quasars:\\
Implications for the Site of \ion{Fe}{2} Line Emission}


\author{Y. Matsuoka\altaffilmark{1}\altaffilmark{2}}
\author{S. Oyabu\altaffilmark{3}}
\author{Y. Tsuzuki\altaffilmark{2}\altaffilmark{4}}
\and
\author{K. Kawara\altaffilmark{1}}

\altaffiltext{1}{Institute of Astronomy, The University of Tokyo, 2-21-1, Osawa,
  Mitaka, Tokyo 181-0015, Japan; matsuoka@ioa.s.u-tokyo.ac.jp.}
\altaffiltext{2}{Visiting astronomer, United Kingdom Infrared Telescope which is operated by the Joint Astronomy Centre on behalf of 
  the U.K. Particle Physics and Astronomy Research Council.}
\altaffiltext{3}{Institute of Space and Astronautical Science, Japan Aerospace Exploration Agency, 3-1-1,
  Yoshinodai, Sagamihara, Kanagawa 229-8510, Japan.}
\altaffiltext{4}{Institute for Cosmic Ray Research, The University of Tokyo, 5-1-5, Kashiwanoha, Kashiwa,
  Chiba, 277-8582, Japan.}


\begin{abstract}
We present results of the near-infrared (IR) spectroscopy of six quasars whose redshifts range from 0.158 to 1.084.
Combined with the satellite ultraviolet data, the relative line strengths of \ion{O}{1} $\lambda$1304, \ion{O}{1} $\lambda$8446,
\ion{O}{1} $\lambda$11287, and the near-IR \ion{Ca}{2} triplet are given.
In addition, the corresponding \ion{O}{1} line strengths measured in normal Seyfert 1s and narrow-line Seyfert 1s are collected 
from the literature.
These lines are thought to emerge from the same gas as do the \ion{Fe}{2} lines, so they are good tracers of the \ion{Fe}{2} emission 
region within a broad emission line region (BELR) in active galactic nuclei (AGNs).
In order to reveal the physical condition within the relevant emission region, we performed photoionized model calculations
and compared them to the observations.
It suggests that a rather dense gas with density $n_{\rm H} \sim 10^{11.5}$ cm$^{-3}$ is present at an
outer portion of the BELR, illuminated by the ionizing radiation corresponding to an ionization parameter
$U \sim 10^{-2.5}$ and is primarily responsible for the observed \ion{O}{1}, \ion{Ca}{2}, and \ion{Fe}{2} lines, based on the 
resemblance of their profiles.
The three \ion{O}{1} lines are proven to be formed through Ly$\beta$ fluorescence and collisional excitation.
We also show that the $\lambda$1304 bump typically observed in AGN spectra consists of the comparable
contributions of \ion{O}{1} and \ion{Si}{2} multiplets, and we discuss the origin of such a strong \ion{Si}{2} emission.
The results are interpreted in the context of the locally optimally emitting cloud (LOC) scenario to find the
plausible gas distribution within the BELR as a function of distance from the central source and density.
\end{abstract}





\section{INTRODUCTION \label{intro}}
The emission lines from singly ionized iron are one of the most prominent features in ultraviolet (UV) to optical spectra of many AGNs.
They are considered to arise from the partly ionized gas present in a BELR.
They dominate the heating and cooling of the emission region through a huge number of electronic transitions
characteristic of the ${\rm Fe}^+$ atom, which results in the \ion{Fe}{2} ``pseudocontinuum'' observed in the AGN spectra.

It is important to understand the physics of the \ion{Fe}{2} line formation for several reasons.
First, since the \ion{Fe}{2} lines carry a large amount of energy away from the BELR, it provides useful information about the energy budget of the emitting gas. 
Second, measurements of the Fe abundance in high-redshift quasars through their \ion{Fe}{2} emission as a function of
cosmic time could verify some cosmological parameters \citep{hamann93, yoshii98}.
Many observations have been devoted to the \ion{Fe}{2} flux measurements in high-redshift 
quasars for this purpose in recent years \citep[e.g.,][]{elston94,kawara96,dietrich02,dietrich03,iwamuro02,iwamuro04,freudling03,maiolino03}.
However, as has been pointed out by several authors \citep[e.g.,][]{verner03,baldwin04,tsuzuki06b}, 
the \ion{Fe}{2} line strengths could be strongly affected by some physical, nonabundance factors of the emitting 
gas, such as density and incident-ionizing radiation flux, so that a priori knowledge of them is definitely needed in order to deduce the Fe abundance.

There have been many studies of \ion{Fe}{2} emission in AGNs.
\citet{netzer83} and \citet{WNW85} modeled the \ion{Fe}{2} emission of quasars using a model ${\rm Fe}^+$ atom with energy levels up to $\sim$10 eV and 
typical BELR parameters in order to compare them with observations.
Their study was followed by, e.g., \citet{sigut98,sigut03} and \citet{verner99,verner03}, who made significant improvements both on the atomic 
data and on the physical processes included.
\citet{baldwin04} calculated the \ion{Fe}{2} and other emission-line strengths in a wide range of gas physical parameters and integrated them to reproduce the 
observed spectra in the context of the LOC \citep{baldwin95} scenario.
However, these studies using the observed shape and/or strength of the \ion{Fe}{2} pseudocontinuum have several serious problems.
From an observational viewpoint, it is extremely difficult to isolate the \ion{Fe}{2} pseudocontinuum from the UV to optical spectrum of AGNs, which is usually made up 
of several continuum components besides emission lines, such as a power-law continuum and a Balmer continuum.
Recently, \citet{tsuzuki06} presented quasar spectra with a wide wavelength coverage,
extending from 1000 to 7300 \AA, thus providing a detailed discussion on this issue.
From a theoretical viewpoint, it is still not confirmed whether the included number of energy levels of the ${\rm Fe}^+$ atom and excitation mechanisms are sufficient to 
reproduce the physical processes occurring in AGNs.
Specifically, it is widely known that none of the current photoionized models can reproduce the observed strong optical \ion{Fe}{2} emitters \citep[e.g.,][]{collin00}
while the recent reverberation mapping results indicate a direct connection between the optical \ion{Fe}{2} emission and the continuum radiation from the central
source \citep{vestergaard05}.

Given the difficulties raised above, one approach to investigating the physical condition within the \ion{Fe}{2} emission region is to use the lines 
emitted by simple atoms in the same region.
\citet{rodriguez02a} found that near-IR \ion{O}{1}, \ion{Ca}{2}, and \ion{Fe}{2} lines
have very similar profiles in the narrow-line Seyfert 1s (NLS1s), indicating that these low-ionization lines are 
arising from the same portion of the BELR.
They also found a good correlation between the near-IR and optical \ion{Fe}{2} emission, which implies a direct link between them in the sense that they are 
produced by the same excitation mechanisms.
The correlation between \ion{O}{1} and \ion{Ca}{2} line widths and between \ion{Ca}{2} and \ion{Fe}{2} line strengths
has also been reported for Seyfert 1s (Sy1s) and NLS1s \citep{persson88}.
Note that they are a natural consequence of similar ionization potentials of the relevant ions, i.e., 16.2 eV for \ion{Fe}{2}, 
13.6 eV for \ion{O}{1}, and 11.9 eV for \ion{Ca}{2}.
%The widths of these lines are significantly narrower than those of Pa$\beta$ and are similar 
%to those of [\ion{S}{3}] $\lambda$9531, the products of the narrow-emission-line region (NELR). 
%It implies that these lines are emitted at the outermost portion of the BELR.
Thus, \ion{O}{1} and \ion{Ca}{2} lines are good tracers of the interesting and important \ion{Fe}{2} line-emission region in AGNs.

The first extensive study of the physical condition within an \ion{O}{1} emission region in AGNs was presented by \citet{grandi80}.
He found that \ion{O}{1} $\lambda$8446 lacks the narrow component that characterizes other permitted lines,
leading to the conclusion that the line is purely a BELR phenomenon.
Such a conclusion is also reported in \citet{riffel06}.
\citet{grandi80} also suggested that \ion{O}{1} $\lambda$8446 is produced by Ly$\beta$ fluorescence, which
was later confirmed by the observation of I Zw 1, the prototype NLS1, by \citet{rudy89}.
\citet{rodriguez02b} compiled the UV and near-IR \ion{O}{1} line strengths of NLS1s (including I Zw 1) 
and of Sy1s, to investigate their flux ratios,
and found that there must be an additional excitation mechanism of \ion{O}{1} $\lambda$8446---besides Ly$\beta$ fluorescence---which 
they concluded is collisional excitation.
\citet{matsuoka05} found that the collisional processes are at work to a similar extent
in quasars and suggested that the gas density in the relevant line emission region is not significantly
different among NLS1s, Sy1s, and quasars.
On the other hand, \citet{ferland89} computed the strengths of various emission lines, including the UV and near-IR
\ion{Ca}{2} lines, and found that a very large column density ($\sim 10^{25}$ cm$^{-2}$) is
needed in order to reproduce the observed relative strengths of these lines. % observed in 25 low-redshift ($z <$ 0.158) AGNs.
However, previous studies have not yet revealed some important physical parameters of the emitting gas quantitatively, mainly due to the lack of 
sufficient data or precise modeling.

Here we present a new near-IR observation of six quasars with redshifts up to 1.084.
Combined with the satellite UV data, the relative strengths of \ion{O}{1} $\lambda$1304, \ion{O}{1} $\lambda$8446, \ion{O}{1} $\lambda11287$, and
the near-IR \ion{Ca}{2} triplet ($\lambda$8498, $\lambda$8542, and $\lambda$8662) are given.
The purpose of the observation is twofold: to give a quantitative insight into the physical condition within the partly
ionized region where \ion{O}{1}, \ion{Ca}{2}, and \ion{Fe}{2} lines are emitted, by comparing the 
observations with precise model calculations;
and to extend the observed \ion{O}{1} and \ion{Ca}{2} samples to higher redshifts than the previous measurements (z $<$ 0.18) and to include quasars, 
which are important in the context of the Fe abundance investigation in high-redshift quasars.
The observations are presented in \S \ref{sec:obs}, and compared model calculations are given in \S \ref{sec:model}.
A discussion of the results and their implications appear in \S \ref{sec:discuss}.
Finally our conclusions are summarized in \S \ref{sec:summary}.

\section{OBSERVATIONS \label{sec:obs}}

\subsection{Target Selection}

Target quasars were selected from a UV spectral atlas of AGNs presented by \citet{evans04}.
The strongest constraint on the target selection comes from their redshifts; \ion{O}{1} $\lambda$8446, \ion{O}{1} $\lambda$11287,
and the near-IR \ion{Ca}{2} triplet should fall within the atmospheric windows.
It significantly reduced a number of possible targets, since these lines are too separated to be observed in the single near-IR band ($J$, $H$, or $K$).
From the quasars satisfying this criterion, we selected those having the highest signal-to-noise ratios (S/Ns) around
\ion{O}{1} $\lambda$1304 in the corresponding UV spectra, which constitute our final target list.
The characteristics of the observed quasars are summarized in Table \ref{sample}.

\subsection{Near-Infrared Observation}

A near-IR observation was carried out at the United Kingdom Infrared Telescope (UKIRT) using the UIST
\citep[UKIRT 1--5 \micron\ Imager Spectrometer;][]{ramsay04} on 2006 February 24.
We adopted the standard UIST observing templates for faint point-source spectroscopy and for point-source imaging.
An observational journal is given in Table \ref{irsp} for the spectroscopy and in Table \ref{irim} for the imaging.
The spectroscopic observation includes a flat, arc, and rationing star for each target, taken right before the target observation.
The rationing stars were selected from the F4 -- F9 type dwarf bright stars with moderately strong \ion{H}{1} recombination lines and weak 
atomic lines, which are well suited for rationing low-resolution spectra.
They were observed at air masses similar to the corresponding targets.
The targets and the rationing stars were observed by nodding the telescope in such a way that they slid +6\arcsec\ and then $-$6\arcsec\ along the slit.
A 0.6\arcsec\ slit yielded an instrumental resolution of $\sim$750 km s$^{-1}$, while the pixel scale is 0.12\arcsec.
The imaging observation consists of dark, flat, and standard stars taken right before the target observation, while the self-flats were used 
for some faint sources.
The standard stars were taken from the UKIRT Faint Standards \citep{hawarden01,legett06} so that they could be observed at air masses similar to the targets.
The targets and the standard stars were observed at five positions on the detector, offset by 20\arcsec\ for the targets and by 10\arcsec\ for the standard stars.
The sky condition was photometric throughout the night, and the average seeing was 0.6\arcsec.

We used the standard method to reduce the data, including removal of dark, flat-fielding, and sky subtraction
using the reduction software IRAF.\footnote{IRAF is distributed by 
the National Optical Astronomy Observatory,
which is operated by the Association of Universities for Research
in Astronomy, Inc., under cooperative agreement with the National
Science Foundation.}
Wavelength calibration was achieved using argon arc-lamp spectra.
In order to calibrate the sensitivity variation along the dispersion axis, the rationing-star spectra---from which the intrinsic \ion{H}{1} recombination 
absorption lines had been removed---were compared with the blackbody spectra with corresponding temperatures, i.e., 6500 K for F4 V type stars, 6250 K for F7 V, 
and 6000 K for F8 V and F9 V.
Finally, photometric calibration was obtained by using the imaging data.
Measured $J$- or $H$-band magnitudes of the quasars are listed in the last two columns of Table \ref{sample}.
Statistical errors of the photometry were less than 0.01 mag for all targets.

We show the reduced spectra along with the sensitivity curves in Figures \ref{plot1} -- \ref{plot2}.
The redshifts measured in the near-IR spectra were in good agreement with those measured in the UV spectra \citep{evans04}.

\subsection{Ultraviolet Data}

Ultraviolet spectra of the quasars were obtained from 1991 January to 1993 July using
the Faint Object Spectrograph (FOS) onboard the {\it Hubble Space Telescope}\footnote{
This publication is based on observations made with the NASA/ESA Hubble Space Telescope, obtained from the data archive at 
the Space Telescope Science Institute, which is operated by the Association of Universities for Research in Astronomy, Inc., 
under NASA contract NAS 5-26555.} ({\it HST}).
In the current work, we used those reduced by \citet{evans04}.
They extensively recalibrated all 
pre-COSTAR (Corrective Optics Space Telescope Axial Replacement) archival FOS UV and optical spectrophotometry
of AGNs with the latest algorithms and calibration data.
If multiple observations of the same source were available, they were combined after being scaled to each other
in such a way that the resultant spectrum would have the highest possible S/N.
The UV spectra around \ion{O}{1} $\lambda$1304 used in this work are shown in Figure \ref{plot_uv}.

\subsection{Emission-Line Measurements}

Emission lines in the UV and the near-IR spectra were manually identified, and their intensities were measured 
by summing up all the flux above the local continuum, which was
%(or continuum plus the wing of the strong emission lines) 
placed by the interpolation method using the observed fluxes in the continuum windows at both sides of the lines.
The continuum windows were carefully selected to not be contaminated by weak emission lines based on the high-S/N
spectra of NLS1s presented by \citet{laor97} and \citet{rodriguez02a}.
%Accordingly, we limited ourselves to deal with the lines having definite continuum windows in the vicinity.
A Voigt profile was fitted to continuum-subtracted spectra by the least $\chi^2$ method to measure line widths for some lines (see below).
As described by \citet{glikman06}, a single-component Voigt profile is a better choice than a multi-component
Gaussian profile when fitting lines with a low S/N.
This is the case for some of the measured lines in our spectra.
When several lines were blended, we fitted a multi-component Voigt profile to the continuum-subtracted spectra
in order to deblend the feature.
Then the total flux was divided into each component, according to the fitted profiles.

Although the $\lambda$1304 bump results from a blending of an \ion{O}{1} triplet at $\lambda$1302,
$\lambda$1305, and $\lambda$1306, and a \ion{Si}{2} doublet at $\lambda$1304 and $\lambda$1309, 
deblending of these components is hopeless, considering the broad line widths of quasars.
Thus, we regarded this feature as a single component in the flux-measurement procedure.
Its constitution is closely discussed in \S \ref{l1304}.
On the other hand, when fitting blended \ion{O}{1} $\lambda$8446 and the \ion{Ca}{2} triplet
($\lambda$8498, $\lambda$8542, and $\lambda$8662), we fixed their relative wavelength centroids and relative intensities
of the \ion{Ca}{2} triplet to the values found in I Zw 1 \citep{rudy00} in order to avoid too many parameters being used 
in the fitting.
This procedure was proven to work very well, since the blueward part of \ion{O}{1} $\lambda$8446 is almost free
from contamination of the \ion{Ca}{2} components.
In some objects, there is an unidentified weak emission line at the blueward part of the blend feature, or the redward part falls
in the atmospheric absorption bands, which limits the available wavelength range used in the fitting and flux measurements.
In these cases, the correction was applied to the flux falling in outside the flux-measured wavelength range according to the 
fitted Voigt profiles.
The fits for the relevant \ion{O}{1} and \ion{Ca}{2} lines in 3C 273 are shown in Figure \ref{fit}.

The measured line fluxes are summarized in Table \ref{lineflux}.
The quoted errors were estimated by the $\Delta \chi^2$ test, which reflect the uncertainties in determining the underlying
continuum level (i.e., the S/N in the continuum windows) and in deblending the blended feature.

\subsection{Variability and Reddening Effects \label{vr}}

It is widely known that quasars are highly variable.
Since the dates of the UV observations are separated from that of the near-IR observation by more 
than 15 yr, a possible variability effect should be carefully examined before comparing the UV and the near-IR spectra.
In fact, the observed quasars show brightness variations in the $J$ or $H$ band by as large as 0.6 mag 
from the dates of the Two Micron All Sky Survey\footnote{
This publication makes use of data products from 2MASS, which is a joint project 
of the University of Massachusetts and the Infrared Processing and Analysis Center/California Institute 
of Technology, funded by the National Aeronautics and Space Administration and the National Science
Foundation.} (2MASS)
observations as shown in Table \ref{sample}. 
However, amounts of the emission-line variabilities corresponding to these continuum variations are rather uncertain.
Monitoring observations of the well-known variable Seyfert NGC 5548 revealed that when the UV 
continuum varied its brightness by about 300\%, the flux variation in \ion{Mg}{2}, one of the low-ionization
lines such as \ion{O}{1} and \ion{Ca}{2}, was only about 10\% \citep{dietrich95}.
On the other hand, \citet{vestergaard05} showed that the variability amplitude of optical \ion{Fe}{2}
emission is 50--75\% of that of H$\beta$, whose variation is comparable to those in the continuum, and UV \ion{Fe}{2}
emission varies with an even larger amplitude. % than H$\beta$.
The response of the individual lines to the continuum variation depends on their formation processes and gas distribution within the BELR 
\citep[e.g.,][]{obrien95}, whose quantitative estimates are far beyond the scope of this paper.
We give notes as to how the possible variability could affect our conclusions in the following sections, when needed.

Another source of errors in comparing the UV and the near-IR spectra is the extinction by dust both in the Galaxy and
in the quasar host galaxies.
We adopted the Galactic extinction values presented by \citet{SFD98}, which are derived from the far-IR observations in two satellite
missions, the {\it Infrared Astronomical Satellite} ({\it IRAS}) and the {\it Cosmic Background Explorer} ({\it COBE}).
Their extinction map has a relatively fine resolution of 6.1\arcmin\ and an accuracy of 16\%.
Corresponding dereddening of the spectra was achieved by using the Galactic extinction curve presented by \citet{pei92}.

On the other hand, much less has been confirmed about the internal reddening of quasars.
\citet{cheng91} found that UV continuum slopes for a wide variety of AGNs were essentially the same to within
their measuring errors, corresponding to a reddening amount of $E_{\rm B-V}$ $\la$ 0.1 mag.
In the optical, \citet{dezotti85} showed that both the broad emission lines and continua of radio-quiet Seyfert galaxies typically 
have a reddening of $E_{\rm B-V}$ = 0.25 mag.
Assuming the Small Magellanic Cloud (SMC) extinction curve \citep{hopkins04} presented by \citet{pei92},
this translates to $E_{\rm 1304-8446}$, the color excess between 1304 and 8446 \AA, of 3.58 mag. 
Hence, the measured UV to near-IR line ratios should be regarded as the lower limits on their intrinsic values,
which could be an order of magnitude or more larger than observed.
As in the case of the variability, the possible changes in our conclusions due to the internal reddening effect are noted in the
following sections.

\section{COMPARISON WITH MODELS \label{sec:model}}

In this section, we aim to investigate the physical condition within the \ion{O}{1} and \ion{Ca}{2} emission region by 
comparing the observed line strengths with model predictions.
It is becoming a common understanding that the BELR is composed of gas with widely distributed physical parameters 
and that each emission line arises from its preferable environment \citep{baldwin95}.
In order to reveal the distribution functions of gas in such a complex system,
%as a function of the distance from the central continuum source and of the gas density, 
it is crucial to constrain which region in the parameter space actually dominates observed emission of the specific lines.

%We emphasize that we concentrate on the observed quantities regarding \ion{Fe}{2}, \ion{O}{1}, and \ion{Ca}{2} lines, 
%which have been revealed to arise from the same portion of the BELR, among the various emission lines observed.
%It makes the single-cloud treatment of the line emission region a very good approximation, which significantly simplifies
%the interpretation of the observed quantities.
%Below we try to determine the dominant region on the parameter space, which should be result of the multiplication of
%the line process efficiency and the distribution function of the gas clouds.

\subsection{Observational Constraints \label{obscst}}

The observed equivalent widths (EWs) of \ion{O}{1} $\lambda$8446 and photon flux ratios of \ion{O}{1} $\lambda$11287 and the near-IR 
\ion{Ca}{2} triplet to \ion{O}{1} $\lambda$8446 are used to constrain the models.
Hereafter, they are expressed as EW (\ion{O}{1} $\lambda$8446), \ion{O}{1} $n$($\lambda$11287)/$n$($\lambda$8446), and
$n$(\ion{Ca}{2})/$n$(\ion{O}{1} $\lambda$8446), respectively.
These emission lines have been revealed to arise from the same portion of the BELR among the various lines observed
(see \S \ref{intro}), which significantly simplifies the interpretation of the observed quantities.
They are summarized in Table \ref{lineratios} ({see the quasar section) for the quasars observed in this work and for the quasar
PG 1116+215 ($z$ = 0.176) studied by \citet{matsuoka05}.
A representative wavelength of 8579 \AA\ was used to convert total flux of the \ion{Ca}{2} triplet to the photon number flux.

The observed values of \ion{O}{1} $n$($\lambda$11287)/$n$($\lambda$8446) are fairly similar from quasar to quasar
except for 3C 232.
The mean value is 0.46 $\pm$ 0.18, where the quoted error reflects a scatter among the quasars.
It should be an excellent indicator of the emitting gas physics, since several important
but uncertain parameters could be nearly canceled out, which include chemical composition and the covering fraction of the line-emitting gas seen 
from the central continuum source.
The reddening effect on the ratio would also be small.
That observed in 3C 232 is significantly different from other quasars, which is apparently due to the extreme weakness of \ion{O}{1} $\lambda$8446.
It is noted that \ion{Ca}{2} is also very weak in 3C 232.
Since there are no distinct features in the sensitivity curve around the relevant wavelength range,
the weakness is not a consequence of the instrumental or the atmospheric absorption effects.
On the other hand, no appropriate models considered below could reproduce such a large value of the \ion{O}{1} $n$($\lambda$11287)/$n$($\lambda$8446) ratio.
Hence, we concluded that some unusual effects concealed a large fraction of emission lines around 8500 \AA\
and decided to exclude this quasar from the sample to be compared with model calculations below.

EW (\ion{O}{1} $\lambda$8446) represents the efficiency of reprocessing the incident continuum radiation into the line emission.
It also depends on the gas covering factor and abundance of the relevant element (oxygen in this case).
Models should reproduce the observed values, EW (\ion{O}{1} $\lambda$8446) $>$ 10 \AA, with reasonable values of these parameters.

The \ion{Ca}{2} triplet is observed to have a strength comparable to \ion{O}{1} $\lambda$8446.
The ratio between them, predicted in model calculations, also depends on the assumed chemical composition (Ca/O in this case) of the emitting gas.
\citet{hamann99} showed that the Mg/O abundance ratio varies from the solar value within a factor of 2 through the course of
galaxy evolution in either the solar neighborhood or the giant elliptical model.
Since Ca and Mg belong to the $\alpha$-elements so that their production rates should not significantly differ from
each other, the Ca/O ratio could also deviate from the solar value by a similar amount.
%Moreover, \citet{ferland89} showed that the near-IR \ion{Ca}{2} emissions are build up beyond the canonical column density of the BELR gas.
%Thus, assumed column density would also affect the predicted line strengths.

\subsection{Model Calculations}

A partial Grotrian diagram of an \ion{O}{1} atom is given in Figure \ref{oienergy}.
In the simplest case, \ion{O}{1} $\lambda$11287, $\lambda$8446, and $\lambda$1304 are formed through Bowen 
resonance-fluorescence by Ly$\beta$, or Ly$\beta$ fluorescence; Ly$\beta$ photons pump up
the ground-term ($2p\ ^3P$) electrons to the excited $3d\ ^3D^0$ level, from which they cascade down to the
ground term through $3p\ ^3P$ and $3s\ ^3S^0$, producing the $\lambda$11287, $\lambda$8446, and $\lambda$1304 line photons.
Note that the pumped $2p\ ^3P$ -- $3d\ ^3D^0$ transition energy corresponds to 1025.77 \AA\ so that
it falls within a Doppler core of Ly$\beta$ (1025.72 \AA) for gas at a temperature of 10$^4$ K.
As a result, the photon flux ratios between these lines should be unity.
However, there could be other excitation mechanisms that alter the ratios, which are collisional excitation, 
continuum fluorescence, and recombination.
\citet{kk81} reported that trapping of $\lambda$1304 becomes important in sustaining the population at $3s\ ^3S^0$ level with the large column density 
typical of the BELR cloud, leading to collisional excitation of the $\lambda$8446 emission.
If $\lambda$8446 also became optically thick, an additional $\lambda$11287 excitation would follow.
Continuum fluorescence of the ground-term ($2p\ ^3P$) electrons could also produce these lines, as well as other lines, such as 
$\lambda$13165 (see Fig. \ref{oienergy}).
This mechanism is known to be an important contributor to the $\lambda$8446 emission observed in some Galactic objects
\citep{grandi75a,grandi75b,rudy91}.
While recombination is another source of the \ion{O}{1} line emissions, its contribution is known to be negligible 
in the BELR gas \citep{grandi80,rodriguez02b}.

We performed model calculations in the framework of 
the photoionized BELR gas using the photoionization code Cloudy, version 06.02 \citep{ferland98}.
The chosen form of the incident continuum shape is a combination of a
$f_{\nu}\propto{\nu}^{\alpha_{\rm uv}} \exp{(-h\nu/kT_{\rm cut})}$
UV bump with an X-ray power law of the form 
$f_{\nu}\propto{\nu}^{\alpha_{\rm x}}$
spanning 13.6 eV to 100 keV.
The UV bump is also cut off in the infrared with a temperature kT$_{\rm IR}$ = 0.01 Ryd,
corresponding to 9.1 \micron.
The UV and X-ray continuum components are combined using a UV to X-ray logarithmic spectral 
slope $\alpha_{\rm ox}$, which is defined by 
$f_{\nu}$(2 kev)/$f_{\nu}$(2500 $\AA$) = 403.3$^{\alpha_{\rm ox}}$.
We adopted a parameter set of [T$_{\rm cut}$, $\alpha_{\rm uv}$, $\alpha_{\rm x}$, $\alpha_{\rm ox}$ =
(1.5 $\times$ 10$^5$) K, $-$0.2, $-$1.8, $-$1.4].
It is essentially the same as those of \citet{tsuzuki06, tsuzuki06b} who derived these parameter values
by combining the results of the thermal accretion disk models \citep{krolik88,binette89,zheng95}
and the observed spectrum of the quasar PG 1626+554 which has typical spectral properties among their 14 quasar samples.

The BELR gas was modeled to have a constant hydrogen density $n_{\rm H}$ (cm$^{-3}$) and exposed to the ionizing continuum
radiation with a photon flux of $\Phi$ (s$^{-1}$ cm$^{-2}$).
The incident-ionizing continuum flux is expressed with an ionization parameter in the calculation
$U$ $\equiv$ $\Phi $/($n_{\rm H}$ $c$), where $c$ is the speed of light, since predicted line strengths tend to be similar along the same values of $U$.
We performed the calculations with the ($n_{\rm H}$, $U$) sets in a range of 10$^{7.0}$ cm$^{-3}$ $\le$ $n_{\rm H}$ $\le$ 
10$^{14.0}$ cm$^{-3}$ and 10$^{-5.0}$ $\le$ $U$ $\le$ 10$^{0.0}$ stepped by 0.5 dex, which should
more than cover the parameter space of the BELR gas.
We began with a BELR gas with $N_{\rm H} = 10^{23}$ cm$^{-2}$ and $v_{\rm turb} = 0$ km s$^{-1}$, where $N_{\rm H}$ is the gas column density and
$v_{\rm turb}$ the microturbulent velocity.
Chemical composition was assumed to be solar.
We hereafter refer to this model as the standard model, or model 1; input parameters are summarized in Table \ref{models}.
%as well as those of other models described below.

In the calculations, the \ion{O}{1} atom was treated as a 6 electron energy-level system including $3s\ ^3S^0$, $4s\ ^3S^0$,
$2p\ ^3P$, $3p\ ^3P$, $4p\ ^3P$, and $3d\ ^3D^0$ levels \citep[see Fig. \ref{oienergy};][]{hazy2}. 
Note that the contribution from recombination is not included in the predicted line strengths, since the
effective recombination coefficients for the \ion{O}{1} lines are not known for densities as large as those in the BELR (G. J. Ferland 2006, private communication).
However, it is expected to be negligible compared to other excitation mechanisms as reported by \citet{grandi80} and \citet{rodriguez02b}.
They showed that the observed strengths of \ion{O}{1} $\lambda$7774, the quintet counterpart to \ion{O}{1} $\lambda$8446,
were less than 10\% of those of 
$\lambda$8446 (see Table \ref{otheroi}), while the ratio between them should be $\lambda$7774/$\lambda$8446 $\sim$ 1.7 in the pure recombination
case based solely on the relative statistical weights \citep{grandi80}.
On the other hand, we found that the ratio should be $\lambda$7774/$\lambda$8446 $\sim$ 7.2 in the pure recombination case
using the effective recombination coefficients in the low-density limit (appropriate for the 
normal Galactic gaseous nebulae) presented by \citet{pequignot91}.
Even if we assumed that the observed $\lambda$7774 was exclusively produced by recombination, its contribution to the $\lambda$8446 is less than 
a few percent.

\subsection{Results}

\subsubsection{The Standard Model}

We show the calculated \ion{O}{1} photon flux ratios $n$($\lambda$11287)/$n$($\lambda$8446) and $n$($\lambda$1304)/$n$($\lambda$8446)
in Figure \ref{oicaii1} as a function of the gas density $n_{\rm H}$ and the ionization parameter $U$.
Note that the incident-ionizing continuum flux $\Phi $ increases toward the direction shown by the arrows in the figure.
%the lines of constant incident ionizing continuum flux, $\Phi$ (= $U$ $n_{\rm_H}$ c) = const., run diagonally from
%top left to bottom right in the figure.
Both ratios are almost unity at the canonical BELR parameters ($n_{\rm H}$, $U$) = (10$^{10}$ cm$^{-3}$, 10$^{-2}$), which refer
to the region where high-ionization lines (such as \ion{C}{4}) are emitted \citep{davidson79}, 
indicating that Ly$\beta$ fluorescence is actually the main formation mechanism of these \ion{O}{1} lines.
As the density becomes large relative to the canonical parameter, both ratios decrease due to the collisional enhancement
of the $\lambda$8446 photons.
On the other hand, continuum fluorescence becomes important as incident continuum flux is significantly decreased, due to a significant 
suppression in Ly$\beta$ formation.
It results in $\lambda$1304 photons not accompanied by those of $\lambda$8446, as well as
$\lambda$8446 not accompanied by $\lambda$11287, so that $n$($\lambda$11287) $<$ $n$($\lambda$8446) $<$ $n$($\lambda$1304).
The results of our observation, \ion{O}{1} $n$($\lambda$11287)/$n$($\lambda$8446) = 0.46 $\pm$ 0.18, clearly indicate
that Ly$\beta$ fluorescence is not the only excitation mechanism of these lines.
The excess $\lambda$8446 photons relative to $\lambda$11287 should be produced through either collisional excitation or continuum 
fluorescence, depending on the physical condition within the emitting gas.

The two excitation mechanisms can be distinguished by their continuum reprocessing efficiency, which converts the incident continuum
radiation into the relevant line emission, through the observed EW (\ion{O}{1} $\lambda$8446).
We plot the calculated values of them in Figure \ref{4chartA} ({\it top right}), assuming a covering factor of 1.0, as well as the 
\ion{O}{1} $n$($\lambda$11287)/$n$($\lambda$8446) ratios for reference ({\it top left}).
Since the covering factor in actual quasars may be much smaller, all the parameter sets satisfying 
EW (\ion{O}{1} $\lambda$8446) $>$ 10 \AA\ are consistent with the observation.
However, it is noted that the covering factor should not be much less than 0.1; in such a situation, the predicted EW (\ion{O}{1} $\lambda$8446)
falls well below those observed on the whole parameter plane.
It is also true in other models described below.
Figure \ref{4chartA} shows that the low density gas in which continuum fluorescence dominates the \ion{O}{1} line
formation predicts \ion{O}{1} $\lambda$8446 at the level of EW (\ion{O}{1} $\lambda$8446) $<$ 1 \AA, which can hardly reproduce
the observation.
On the other hand, there is a large area consistent with both the observed \ion{O}{1} $n$($\lambda$11287)/$n$($\lambda$8446) ratios 
and EW (\ion{O}{1} $\lambda$8446) at the high-density side where collisional excitation is at work.
Thus, we concluded that the observed \ion{O}{1} lines are produced by a combination of Ly$\beta$ fluorescence and collisional excitation.

We show the calculated $n$(\ion{Ca}{2})/$n$(\ion{O}{1} $\lambda$8446) ratios in Figure \ref{4chartA} ({\it bottom left}) to further
constrain the model parameters.
The fact that these two emissions are observed to have comparable strengths, combined with the results from the \ion{O}{1} line strengths,
finely restricts the gas density to be around $n_{\rm H}$ = 10$^{11.5}$ cm$^{-3}$.
However, the model predictions could vary within a factor of 2, depending on the assumed chemical composition as described in \S \ref{obscst}.
It also depends on the assumed column density $N_{\rm H}$ and increases by up to a factor of 3 when it is changed from 
$N_{\rm H} = 10^{23}$ to $10^{25}$ cm$^{-2}$, as shown in Figure \ref{cloud_str}.
Thus, we required that the predicted \ion{Ca}{2}/\ion{O}{1} $\lambda$8446 ratio should fall in the range 0.1 $<$ $n$(\ion{Ca}{2}) /
$n$(\ion{O}{1} $\lambda$8446) $<$ 5.0.
This constraint still effectively excludes the high-density ($n_{\rm H} >$ 10$^{12.0}$ cm$^{-3}$) and
low ionization parameter ($U <$ 10$^{-2.5}$) models in which the \ion{Ca}{2} emission is significantly overpredicted.

The top left, top right, and bottom left panels of Figure \ref{4chartA} are shown in a single plot in the 
bottom right panel of the same figure.
The constraints from the observed values of \ion{O}{1} $n$($\lambda$11287)/$n$($\lambda$8446), EW (\ion{O}{1} $\lambda$8446),
and $n$(\ion{Ca}{2})/$n$(\ion{O}{1} $\lambda$8446) are shown by thick solid lines, thin solid lines, and the shaded area, respectively.
The models consistent with all the constraints are marked with diamonds, which are $n_{\rm H}$ = 10$^{11.5}$ cm$^{-3}$, and 
$U$ = 10$^{-2.5}$ and 10$^{-3.0}$.
However, since the covering factors in actual quasars are known to be $\sim$0.1, we favor the models predicting
EW (\ion{O}{1} $\lambda$8446) $>$ 100 \AA\ despite the formal constraint of $>$ 10 \AA.
Thus, we concluded that the best-fit parameter set to the observations is ($n_{\rm H}$, $U$) = (10$^{11.5}$ cm$^{-3}$, 10$^{-3.0}$) in the standard model.
The predicted line strengths in this specific model are summarized in Table \ref{bfmodels}.

%Next we set gas column density to $N_{\rm H} = 10^{25}$ cm$^{-2}$ while other input parameters are identical to the standard 
%model with ($n_{\rm H}$, $U$) = ($10^{11.5}$ cm$^{-3}$, $10^{-3.0}$), to investigate the effect of changing $N_{\rm H}$ on the predicted line strengths.
%Results are shown in Figure \ref{cloud_str}, in which cumulative ratios of \ion{O}{1} $n$($\lambda$11287)/$n$($\lambda$8446), 
%\ion{O}{1} $n$($\lambda$1304)/$n$($\lambda$8446), and $n$(\ion{Ca}{2})/$n$(\ion{O}{1} $\lambda$8446) as well as cumulative
%flux of \ion{O}{1} $\lambda$8446 are plotted against gas column density measured from the illuminated surface.
Next we investigated the line strengths as a function of gas column density in the range $N_{\rm H} = 10^{17} - 10^{25}$ cm$^{-2}$.
A calculation was performed for the gas with $N_{\rm H} = 10^{25}$ cm$^{-2}$, while other input parameters are identical to the standard 
model with ($n_{\rm H}$, $U$) = ($10^{11.5}$ cm$^{-3}$, $10^{-3.0}$), and the cumulative ratios of \ion{O}{1} $n$($\lambda$11287)/$n$($\lambda$8446), 
\ion{O}{1} $n$($\lambda$1304)/$n$($\lambda$8446), and $n$(\ion{Ca}{2})/$n$(\ion{O}{1} $\lambda$8446), as well as the cumulative
flux of \ion{O}{1} $\lambda$8446, are plotted against the column density measured from the illuminated surface in Figure \ref{cloud_str}.
The \ion{O}{1} $\lambda$8446 emission is mainly built up before the column density $N_{\rm H} = 10^{23}$ cm$^{-2}$ is reached, and the
deeper region contributes only $\sim$10\% to the emergent emission.
Correspondingly, the two \ion{O}{1} flux ratios stay nearly constant between $N_{\rm H} = 10^{22}$ and $10^{25}$ cm$^{-2}$.
On the other hand, the $n$(\ion{Ca}{2})/$n$(\ion{O}{1} $\lambda$8446) ratio becomes $\sim$ 2 times larger when the column density
is increased from $N_{\rm H}$ = 10$^{23}$ to 10$^{25}$ cm$^{-2}$.
This effect was first pointed out by \citet{ferland89} and is already taken into account in the above arguments.
Hence, we concluded that the assumed column density does not significantly change the results from the standard model unless it is much smaller 
than $\sim$10$^{22}$ cm$^{-2}$.
Note that the \ion{O}{1} line formation would be significantly suppressed in such a thin gas (see Fig. \ref{cloud_str}) so that the EWs
fall well below the observed values.

\subsubsection{Other Models: Microturbulence and Incident Continuum\label{sec:mi}}

Some BELR models indicate a presence of the microturbulence within the gas, which includes a large velocity
gradient in a wind \citep[e.g.,][]{konigl94} and nondissipative magnetohydrodynamic waves in magnetically confined clouds \citep{rees87,bf00}.
We performed another set of calculations with the microturbulence included, whose velocity is $v_{\rm turb}$.
It is known to have a large influence on the \ion{Fe}{2} line strengths through a combined effect of the increased continuum fluorescence rate and 
the decreased line optical depth \citep{netzer83,verner99}.
Using a strong dependency, \citet{verner03} showed that $v_{\rm turb}$ = 5 -- 10 km s$^{-1}$ gives the most plausible values
to account for the observed \ion{Fe}{2} and \ion{Mg}{2} emissions.
On the other hand, \citet{baldwin04} reported that the photoionized BELR models cannot reproduce both the observed
shape and EW of the \ion{Fe}{2} UV bump unless the considerable amount of microturbulent velocity corresponding to
$v_{\rm turb} \ge$ 100 km s$^{-1}$ is included.
Thus, we considered two models with the microturbulent velocity set to $v_{\rm turb}$ = 10 km s$^{-1}$ (model 2) and 100 km s$^{-1}$
(model 3).
The input parameters of these models are summarized in Table \ref{models}.

The results in models 2 and 3 are rather similar to the standard model (Fig. \ref{4chartA}), while the whole patterns of contour 
are shifted a little toward a high-density regime as the microturbulent velocity is increased.
It is thought to be due to the decreased optical depth in \ion{O}{1} $\lambda$1304, which then reduces the electron population in
the $3s\ ^3S^0$ level and thus the collisional excitation rate of \ion{O}{1} $\lambda$8446.
As a result, it has to be compensated by the increased particle density to reproduce the observed degree of collision rate.
It is also noted that the predicted EWs of \ion{O}{1} lines are moderately decreased on the whole ($n_{\rm H}$, $U$) parameter plane 
in model 3.
Since the difference in the wavelengths of Ly$\beta$ (1025.72 \AA) and the pumped \ion{O}{1} transition $2p\ ^3P$ -- $3d\ ^3D^0$
(1025.77 \AA) corresponds to the velocity of 15 km s$^{-1}$, the Ly$\beta$ fluorescence efficiency is reduced in such a turbulent gas.
The best-fit ($n_{\rm H}$, $U$) parameters are determined in the same manner as in the standard model and listed in Table \ref{bfmodels}, which are 
($n_{\rm H}$, $U$) = ($10^{11.5}$ cm$^{-3}$, $10^{-2.5}$) in model 2 and ($10^{12.0}$ cm$^{-3}$, $10^{-2.5}$) in model 3.
The predicted line strengths in these specific models are also listed in the table.

Finally, we changed the incident continuum shape to be much harder as suggested by \citet{korista97}, which has the form
($T_{\rm cut}$, $\alpha_{\rm uv}$, $\alpha_{\rm x}$, $\alpha_{\rm ox}$) = (10$^6$ K, $-$0.5, $-$1.0, $-$1.4).
It was also adopted by \citet{baldwin04}.
The microturbulent velocity is set to $v_{\rm turb}$ = 100 km s$^{-1}$, which makes this model identical to model 8 of \citet{baldwin04}.
Model 8 was considered to reproduce both the observed shape and EW of the \ion{Fe}{2} UV bump observed in AGNs with
strong \ion{Fe}{2} emission, and was proven to be one of the successful models (see \S \ref{role} for more discussions).
This is our model 4, whose input parameters are summarized in Table \ref{models}.
Although the incident continuum shape is dramatically changed, the results are nearly identical to those in model 3, except that the predicted
\ion{O}{1} EWs are further decreased.
The best-fit parameters are identical to model 3, ($n_{\rm H}$, $U$) = ($10^{12.0}$ cm$^{-3}$, $10^{-2.5}$).
%However, this model has an advantage that it can reproduce both the observed shape and equivalent width of UV \ion{Fe}{2} bump
%\citep[][; see below for more discussions]{baldwin04}.
They are listed in Table \ref{bfmodels} along with the predicted line strengths.

\section{DISCUSSION \label{sec:discuss}}

\subsection{Excitation Mechanisms of the \ion{O}{1} Lines}

Our results indicate that the \ion{O}{1} emission observed in quasars is formed through Ly$\beta$ fluorescence and collisional
excitation in gas with a density of $n_{\rm H} \sim$ 10$^{11.5}$ cm$^{-3}$, illuminated by the ionizing radiation corresponding to
the ionization parameter $U \sim$ 10$^{-2.5}$.
It can be tested by investigating the observed strengths of other \ion{O}{1} lines, which should or should not be accompanied by 
the strongest three lines, $\lambda$1304, $\lambda$8446, and $\lambda$11287, in each excitation mechanism.
We list the observed flux ratios of these lines measured to date in Table \ref{otheroi}.
Theoretical predictions of the ratios for the four possible excitation mechanisms, i.e., Ly$\beta$ fluorescence, collisional excitation, 
continuum fluorescence, and recombination, are collected from the literature and also listed. 
In the case of pure Ly$\beta$ fluorescence, all the ratios listed in the table should be zero; actually, this is satisfied in
most objects with one exception, 1H 1934$-$063.
The Ly$\beta$ fluorescence plus collisional excitation from the $3s\ ^3S^0$ level, which our results indicate, would also
predict all the ratios in the table to be nearly zero because the collisional excitation would predominantly produce the $\lambda$8446 line
photons against other lines listed.

The detection of $\lambda$7774 in 1H 1934$-$063 led \citet{rodriguez02b} to conclude that collisional excitation plays an important role in 
the \ion{O}{1} line formation in NLS1s.
They stated that the observed $\lambda$7774/$\lambda$8446 ratio is $\sim$0.2 after the removal of Ly$\beta$ fluorescence contribution, 
which is consistent with the theoretical prediction of pure collisional excitation given by \citet{grandi80} 
($\lambda$7774/$\lambda$8446 $\sim$ 0.3; the ``collisional excitation'' row in Table \ref{otheroi}).
However, the prediction was derived from the collisional cross sections only from the ground term ($2p\ ^3P$).
It should be an oversimplification because the collisional excitation from the excited $3s\ ^3S^0$ level, which we have shown
is a large contributor to the $\lambda$8446 emission, is not included.
Then the observed strength of $\lambda$7774 in 1H 1934$-$063, as large as 20\% of $\lambda$8446 (after Ly$\beta$ fluorescence 
contribution was subtracted), should be attributed to other excitation mechanisms.
Since another line, $\lambda$7990, is also detected in similar strength in 1H 1934$-$063, recombination might be working 
to some extent and producing these emissions, as \citet{rodriguez02b} recognized.
An absence of the fourth mechanism, continuum fluorescence, is evident from the observed weakness of $\lambda$13165; it is detected in none of 
the observed samples, while it should have a strength comparable to $\lambda$11287 in the pure continuum fluorescence case \citep{grandi80}.

Yet another mechanism of the \ion{O}{1} line formation was suggested by \citet{grandi83}, in which the $3s\ ^3S^0$ level 
could decay to the metastable terms of the ground configuration via the semiforbidden lines
$\lambda$1641 and $\lambda$2324 (see Fig. \ref{oienergy}).
He suggested that fully half of the $\lambda$1304 photons would be destroyed by this mechanism in the typical BELR cloud.
However, \citet{laor97} reported that they found no strong line at 1641 \AA\ in the spectrum of
I Zw 1 and that such a line cannot add more than $\sim$30\% to the observed $\lambda$1304 flux.

\subsection{Constitution of the $\lambda$1304 Bump \label{l1304}}

The $\lambda$1304 bump typically seen in AGN spectra is the result from a blending of an \ion{O}{1} triplet 
at $\lambda$1302.17, $\lambda$1304.86, and $\lambda$1306.03, with a mean laboratory wavelength of 1303.50 \AA\ 
\citep{verner96, constantin02}
and a \ion{Si}{2} doublet at $\lambda$1304.37 and $\lambda$1309.27 (mean laboratory wavelength 1307.63 \AA, hereafter referred to as the $\lambda$1308 doublet).
This blending is very strong in most quasars, preventing a reliable estimate of relative contributions of \ion{O}{1} 
and \ion{Si}{2} to the blend.
Accordingly, deblending of the bump has been tried mostly for NLS1s and quasars having relatively narrow broad emission line profiles.
\citet{baldwin96} performed the method to fit the synthetic spectra constructed separately for each quasar to deblend the bump.
Although they could not firmly determine the strengths of both the \ion{O}{1} and \ion{Si}{2} components for any single quasar, the fitting
results show that the \ion{O}{1} fractions in the bump vary in the range from $<$30\% to $>$60\%.
More plausible measurements were achieved for I Zw 1 by \citet{laor97}.
The \ion{Si}{2} $\lambda$1309.27 component is clearly detected in I Zw 1, which enables the reliable estimate of the \ion{O}{1} and \ion{Si}{2} fractions 
in the blend.
Their template fit shows that 50\% (56\%) of the bump flux is due to \ion{O}{1} in the optically thick (thin) \ion{Si}{2} doublet case.
\citet{rodriguez02b} applied the same method to three NLS1s and found that the average portion of the \ion{O}{1} flux is 75\%.

Now we investigate the composition of the $\lambda$1304 bump from the standpoint of the photoionized model calculations.
We are particularly interested in whether the dominant contributor to the bump, i.e., either \ion{O}{1} or \ion{Si}{2}, 
actually varies from quasar to quasar as shown by \citet{baldwin96}.
We plot the \ion{O}{1} $n$($\lambda$1304)/$n$($\lambda$8446) versus $n$($\lambda$11287)/$n$($\lambda$8446) relation in Figure \ref{oivsoi},
which is calculated in models 1 -- 4 with the ($n_{\rm H}$, $U$) parameters around the best-fit values with which the \ion{O}{1} line formation is
dominated by Ly$\beta$ fluorescence and collisional excitation, as revealed above.
The range of the observed $n$($\lambda$11287)/$n$($\lambda$8446) value is shown by two dotted lines.
In order to obtain a rough estimate independent of the selected model, we regard the predicted $n$($\lambda$1304)/$n$($\lambda$8446)
values enclosed by two dashed lines in the figure, drawn to include most of the points in the observed $n$($\lambda$11287)/$n$($\lambda$8446) range,
as the allowed values against $n$($\lambda$11287)/$n$($\lambda$8446).
These lines are expressed as
\begin{equation}
n(\lambda1304)/n(\lambda8446) \ =\ 1.15\ \times\ n(\lambda11287)/n(\lambda8446)\ -\ (0.20 \pm 0.05).
\end{equation}
Then the observed $n$($\lambda$11287)/$n$($\lambda$8446) ratios and the bump (\ion{O}{1} + \ion{Si}{2} $\lambda$1304) flux give the
estimates of the \ion{O}{1} fraction in the bump; the results are listed in Table \ref{oi1304}.
It shows that the \ion{O}{1} fractions actually vary from quasar to quasar by a significant amount, ranging from
$\la$20\% in QSO J1139$-$1350 to $\ga$60\% in 3C 273 and QSO B0850$+$440.
While it was hoped that \ion{O}{1} $\lambda$1304/$\lambda$8446 ratio would be a reliable reddening indicator of AGNs
as long as its intrinsic value was accurately determined \citep[e.g.,][]{rodriguez02b},
our results show that it is more difficult than had been expected without accurate line deblending or precise model calculations.

The derived fluxes of the \ion{Si}{2} $\lambda$1308 doublet are comparable or even a few times larger than those of 
\ion{O}{1} $\lambda$1304.
However, it cannot be reproduced in our model calculations; the equivalent width of \ion{Si}{2} $\lambda$1308 is at most 3 \AA\ 
on the whole ($n_{\rm H}$, $U$) parameter plane in all models while EW (\ion{O}{1} $\lambda$1304) $\sim$
100 \AA\ (10 \AA\ if the covering factor is set to 0.1) with the best-fit parameters.
Changing the assumed column density does not settle the problem, in which the \ion{Si}{2} $\lambda$1308 flux is enhanced
only by $\sim$20\% when $N_{\rm H}$ is increased from 10$^{23}$ to 10$^{25}$ cm$^{-2}$.
A possible explanation comes from the LOC scenario.
We plot the distribution of EW (\ion{Si}{2} $\lambda$1308) as a function of the gas density $n_{\rm H}$ and the ionization parameter $U$ in the standard 
model in Figure \ref{ewsi}, which shows that the \ion{Si}{2}
$\lambda$1308 emission arises from the vast area on the parameter plane in similar strengths, while \ion{O}{1} emission arises from the relatively
limited region (Fig. \ref{4chartA}, {\it top right}).
Thus, the LOC-type integration over the parameter plane could significantly strengthen the predicted \ion{Si}{2} emission 
relative to \ion{O}{1}, which could account for the observations.
An alternative way to interpret the strong \ion{Si}{2} $\lambda$1308 emission is an unusually large microturbulent velocity present
in the emitting gas, as suggested by \citet{bottorff00}.
According to their calculations, \ion{Si}{2} $\lambda$1308 is one of the most enhanced lines by the increased micoruturbulent velocity.
They showed that very large values of the microturbulent velocity are required in order to reproduce the \ion{Si}{2} line strengths reported by
\citet{baldwin96}, which correspond to $v_{\rm turb} >$ 1000 km s$^{-1}$.
\ion{Si}{2} $\lambda$1308 is enhanced by more than 2 orders of magnitude in such a highly turbulent gas compared to the no
microturbulence case, while \ion{O}{1} $\lambda$1304 is reduced by more than 80\%  due to the significant suppression in the Ly$\beta$ 
fluorescence rate.

Still other mechanisms of the \ion{Si}{2} $\lambda$1308 enhancement are discussed in detail in \citet[][Appendix C]{baldwin96}.
They suggested that dielectric recombination or charge transfer might produce significant emission in the highly excited \ion{Si}{2} lines, 
including $\lambda$1308, in order to account for their observed strengths relative to the lowest line, $\lambda$1814, in one of the gas
components (dubbed ``component A'') in the quasar Q0207$-$398.
Note that component A was found to have an extremely high ($n_{\rm H} \ga 10^{12.5}$ cm$^{-3}$) gas density, which they stated
makes the above processes competitive through the thermalization of all resonance lines.

Finally, it should be noted that the intrinsic values of \ion{O}{1} + \ion{Si}{2} $\lambda$1304/\ion{O}{1} $\lambda$8446 ratio
could be larger than measured by more than an order of magnitude due to the internal reddening effect.
Then the \ion{O}{1} fractions in the $\lambda$1304 bump would become even smaller than derived here, corresponding to the applied amounts of
dereddening.
An additional but probably smaller uncertainty comes from the possible variability event of the emission lines (see \S \ref{vr}).

\subsection{Situation in Other Type 1 AGNs}

Observations of the \ion{O}{1} flux ratio between $\lambda$8446 and $\lambda$11287 in Sy1s and
NLS1s have been reported to date \citep{laor97, rudy00, rodriguez02b}, which are listed in Table \ref{lineratios}
(see the Sy1 and NLS1 sections).
The ratio of 0.55 $\pm$ 0.08 in the Sy1 is very similar to those of quasars, which may indicate similar emitting gas 
properties in the BELR of both types of AGNs.
However, a larger sample of Sy1s is obviously needed in order to make a significant comparison.

On the other hand, the mean \ion{O}{1} $n$($\lambda$11287)/$n$($\lambda$8446) ratio observed in NLS1s is 0.73 $\pm$ 0.20,
which is marginally higher than those of quasars and Sy1s (although they agree within the 1 $\sigma$ deviation).
This situation is illustrated well in Figure \ref{otheragn}, in which we plot the number distributions of the ratio
separately for quasars plus Sy1s and for NLS1s.
It is direct evidence that radiation-related processes are more efficient over collisional processes in the BELR gas in NLS1s 
than in quasars and Sy1s.
A possible explanation might come from the general idea of the NLS1 containing less massive black holes with higher mass
accretion rates, radiating at close to Eddington luminosity \citep{pounds95,laor97a,mineshige00}, than normal Seyfert galaxies.
Consequently, the line-emitting gas might be exposed to stronger continuum radiation so that the radiation-related processes become predominant.
However, in the LOC scenario, the line-emitting region simply moves farther out when the continuum flux is increased at a given radius,
so that the incident continuum flux (presumably the ionization parameter) in the region is kept nearly constant.
Such a situation is actually confirmed by the reverberation mapping results \citep{peterson02}.
The difference in the ionizing continuum shape would also not be the origin of the \ion{O}{1} $n$($\lambda$11287)/$n$($\lambda$8446) difference as we see
in \S \ref{sec:mi}; in fact, the \ion{O}{1} line ratios predicted in models 3 and 4 are quite similar to each other when the gas density and the ionization parameter
are fixed (see Table \ref{bfmodels}).

Alternatively, the somewhat smaller gas density in the BELR of NLS1 gives a natural explanation for the decreased collisional excitation rate.
In this case, the difference in the mean \ion{O}{1} $n$($\lambda$11287)/$n$($\lambda$8446) ratio between quasars plus Sy1s and the NLS1s corresponds to
a few $\times$ 0.1 dex smaller gas density in NLS1s, as seen in Figure \ref{4chartA} ({\it top left}; around the best-fit parameters
log $n_{\rm H} \sim 11.5$ and log $U \sim -3.0$).

\subsection{Role of the Relevant Gas in the BELR \label{role}}

Our results suggest that a rather high ($n_{\rm H} \sim$ 10$^{11.5}$ cm$^{-3}$) density gas is present in the BELR at the region illuminated
by the continuum radiation corresponding to the ionization parameter $U \sim$ 10$^{-2.5}$, producing a dominant part of the observed
\ion{O}{1} and \ion{Ca}{2} emissions.
The gas is expected to be located at an outer portion of the BELR.
It is a natural consequence of the low ionization potentials of the relevant ions relative to other typical emission lines observed 
in AGNs and is also indicated by the observed line widths.
We list the full widths at half maximum (FWHMs) of the observed \ion{C}{4} $\lambda$1549 and \ion{O}{1} $\lambda$11287 line profiles in Table
\ref{fwhm} as the representatives of the high- and low-ionization lines.
These two lines were chosen because they are relatively isolated from other lines and are strong enough to allow reliable measurements.
The table shows that the line widths of \ion{O}{1} $\lambda$11287 are 0.3 -- 0.5 times narrower than those of \ion{C}{4} $\lambda$1549.
Assuming that the BELR kinematics is Keplerian so that $V_{\rm FWHM} \propto r^{-1/2}$, where $V_{\rm FWHM}$ is
the emission-line width and $r$ is the distance between the line-emitting gas and the central source \citep[e.g.,][]{peterson99,peterson00},
the \ion{O}{1} emitting gas seems to be orbiting at a radius several times larger than that of \ion{C}{4} $\lambda$1549.
The reverberation mapping results for the low-ionization lines such as \ion{Mg}{2}, \ion{Fe}{2}, and H$\beta$ 
\citep{dietrich95,peterson99,vestergaard05} also support this picture, although the results for the relevant \ion{O}{1} and \ion{Ca}{2} 
lines have not been obtained to date.

In order to check the consistency of our results with the observations of other emission lines, we list the calculated EWs of typical UV to optical 
lines seen in AGNs in Table \ref{otherlines}.
The calculation was performed with the best-fit ($n_{\rm H}$, $U$) parameters to the \ion{O}{1} and \ion{Ca}{2} observations and the covering factor of 0.1 
in the standard model, while similar results were obtained in other models.
We also list the EWs measured in a Large Bright Quasar Survey (LBQS) composite spectrum \citep{francis91}
for the compared observation.
The composite spectrum was constructed from 688 quasars (absolute magnitudes $M_{\rm B_J}$ $\le$ $-$21.5) and 30 AGNs
(absolute magnitudes $-21.5 \le M_{\rm B_J} \le -20.5$) showing no strong evidence for the presence of broad absorption line troughs 
\citep[see][for details of the observing procedures]{foltz87,morris91}.
The table shows that the \ion{O}{1} and \ion{Ca}{2} emitting gas exclusively contributes to the permitted low-ionization lines 
such as \ion{Mg}{2}, UV \ion{Fe}{2}, and H$\beta$, as we expected.
On the other hand, a main part of high-ionization lines such as \ion{Si}{4} + \ion{O}{6}] $\lambda$1400 and
\ion{C}{4} $\lambda$1549 is not produced in this gas; these lines should be formed in an inner portion of the BELR where gas is illuminated
by stronger ionizing radiation.
The forbidden line [\ion{O}{3}] $\lambda$5007 is also deficient in the emitted spectrum from the gas, which is consistent with the general picture
that the line is formed in a much lower density environment in the NELR.

The shape of the \ion{Fe}{2} UV bump is correctly reproduced in models 3 and 4, in which $v_{\rm turb}$ is set to 100 km s$^{-1}$ as found
by \citet{baldwin04}.
In fact, the best-fit parameters in these models, ($n_{\rm H}$, $U$) = (10$^{12.0}$ cm$^{-3}$, 10$^{-2.5}$), lie on the parameter
region consistent with the observed EW and ``spike/gap'' ratio they defined (see their Fig. 7; our models correspond to 
log $n_{\rm H}$ = 12.0 and log $\Phi_{\rm H}$ = 20.0).
Models 1 (the standard model) and 2 cannot reproduce the observed spike/gap ratio; the bump is dominated by the strongest resonance lines,
making the predicted ratio much larger than observed.
On the other hand, none of the models can reproduce the strong optical \ion{Fe}{2} emitters.
It has long been a challenge to the photoionized models \citep[e.g.,][]{collin00} and was recently reviewed by \citet{baldwin04}, who stated 
that the bulk of the optical \ion{Fe}{2} emission might come from a region different from that which produces the UV \ion{Fe}{2} emission, 
such as collisionally ionized gas.
We will return to this issue in a forthcoming paper \citep{tsuzuki06b}.

\subsection{Interpretation of the results in the LOC Scenario}

In the context of the LOC scenario, an observed spectrum is the sum of emission from the gas
at a range of different distances from the central source and also within a range of different densities.
Hence the observed emission line flux should be expressed as
\begin{equation}
F_{\rm line} \propto \int\int r^2\ F(r, n_{\rm H})\ f(r)\ g(n_{\rm H})\ dn_{\rm H}\ dr,
\end{equation}
where $F(r, n_{\rm H})$ is the emission-line flux of a single cloud at radius $r$ from the central source and with density $n_{\rm H}$, while 
$f(r)$ and $g(n_{\rm H})$ are distribution functions of the gas covering fraction and density, respectively.

In this scenario, the best-fit ($n_{\rm H}$, $U$) parameters we derived above correspond to the model producing a dominant 
part of the observed \ion{O}{1} and \ion{Ca}{2} emission.
We plot the quantity
\begin{equation}
{\rm DF}(r, n_{\rm H}) \equiv r^2\ F(r, n_{\rm H})\ f(r)\ g(n_{\rm H})\ n_{\rm H}\ r,
\end{equation}
for \ion{O}{1} $\lambda$8446 in Figure \ref{loc} to see which region on the parameter plane actually dominates the emission.
The covering fraction and the gas-density distributions are assumed to be [$f(r) \propto\ r^0,\ r^{-1},\ r^{-2}$] and
[$g({n_{\rm H}}) \propto n_{\rm H}^0,\ n_{\rm H}^{-1},\ n_{\rm H}^{-2}$], respectively.
With $g({n_{\rm H}}) \propto n_{\rm H}^0$ or $n_{\rm H}^{-2}$ density distributions, the \ion{O}{1} $\lambda$8446 emission
is dominated by much higher or lower density gas than the best-fit model.
On the other hand, an $f(r) \propto r^0$ type covering fraction results in dominant emission from the gas that is much farther from the central source
(i.e., illuminated by much weaker ionizing radiation) than the best-fit model, while $f(r) \propto r^{-2}$ predicts the opposite end.
Thus, the functions of $f(r) \propto r^{-1}$ and $g({n_{\rm H}}) \propto n_{\rm H}^{-1}$ represent the most plausible distributions of the BELR gas.
Note that it is also true in other models (models 2 -- 4).
It is a consequence of the fact that any of the assumed ($n_{\rm H}$, $U$) parameters do not predict the EW
of \ion{O}{1} lines much greater than observed, i.e., EW (\ion{O}{1} $\lambda$8446) $\gg$ 10 \AA, with a reasonable
covering factor (see Fig. \ref{4chartA}, {\it top right}), so the best-fit parameters are chosen near the EW (\ion{O}{1} $\lambda$8446)
maximum on the parameter plane.
Since the emission-line flux of a single cloud can be written as $F(r, n_{\rm H})$ = (EW)$F_{\rm continuum} \propto$ (EW)$r^{-2}$,
with $f(r)\ \propto\ r^{-1}$ and $g({n_{\rm H}}) \propto n_{\rm H}^{-1}$ distributions, DF becomes
\begin{equation}
{\rm DF}(r, n_{\rm H}) \propto r^2\ [({\rm EW})r^{-2}]\ r^{-1}\ n_{\rm H}^{-1}\ n_{\rm H}\ r \propto {\rm EW}.
\end{equation}
It means that the DF distribution is identical to those of EW on the parameter plane so that the best-fit ($n_{\rm H}$, $U$) 
parameters chosen near the EW maximum should also be near the DF maximum.
On the other hand, other shapes of $f(r)$ and $g({n_{\rm H}})$ functions predict the DF maxima very differently from that of EW, i.e.,
very differently from the best-fit ($n_{\rm H}$, $U$) parameters.
Thus, our results point to the gas distribution of $f(r) \propto r^{-1}$ and $g({n_{\rm H}}) \propto n_{\rm H}^{-1}$ forms in the LOC model to a
first-order approximation.
%These distribution functions are consistent with the standard LOC model described by \citet{baldwin95}.

On the other hand, the DF maximum for the \ion{Ca}{2} lines occurs at a much lower value of $U$ than the best-fit ($n_{\rm H}$, $U$) parameters 
with $f(r) \propto r^{-1}$ and $g({n_{\rm H}}) \propto n_{\rm H}^{-1}$ distribution functions, as shown in Figure \ref{loc_ca}.
It implies that the gas distribution should have a cutoff somewhere around $U \sim$ 10$^{-3.5}$ at $n_{\rm H} = 10^{11.0}-10^{12.0}$ cm$^{-3}$,
which should originate from the physical nature of the BELR.
In fact, \citet{baldwin95} constrained a range of the LOC integration over the parameter space to $\Phi > 10^{18}$ s$^{-1}$ cm$^{-2}$,
since gas at a larger distance than this limit will form graphite grains \citep{sanders89} so that the line emission is heavily suppressed \citep{netzer93}.
This limit corresponds to $U \ga 10^{-3.5}$ at $n_{\rm H} = 10^{11.5}$ cm$^{-3}$, which is consistent with the above expectation.
%Note that the distribution of $f(r)$ = const., as well as $f(r) \propto r^{-1}$, is also allowed with the introduced cutoff (see Figure \ref{loc}).

\section{SUMMARY \label{sec:summary}}

We present results of the near-IR spectroscopy of six quasars with redshifts ranging from 0.158 to 1.084.
Combined with the UV data taken with the {\it HST} FOS, we give the relative strengths of \ion{O}{1}
$\lambda$1304, \ion{O}{1} $\lambda$8446, \ion{O}{1} $\lambda$11287, and the \ion{Ca}{2} triplet ($\lambda$8498, $\lambda$8542, and $\lambda$8662).
%In addition, these lines previously measured in quasars, Seyfert 1s, and NLS1s were collected from the literature and 
%included in the sample.
Detailed model calculations were performed in the framework of the photoionized BELR gas for a wide range of physical parameters and compared to 
the observations, which led us to the following conclusions.

\noindent 1.\ A rather dense gas with density $n_{\rm H} \sim 10^{11.5}$ cm$^{-3}$ is present at an outer portion of the BELR,
whose distance from the central source corresponds to the ionization parameter $U \sim 10^{-2.5}$.
The gas is primarily responsible for the observed \ion{O}{1}, \ion{Ca}{2}, and \ion{Fe}{2} lines based on the resemblance of their profiles.

\noindent 2.\ The strongest \ion{O}{1} lines observed in typical AGN spectra, i.e., $\lambda$1304, $\lambda$8446, and $\lambda$11287,
are formed through Ly$\beta$ fluorescence and collisional excitation.
It is consistent with the absence or weakness of other \ion{O}{1} lines such as $\lambda$7774 and $\lambda$13165.

\noindent 3.\ The fractions of the \ion{O}{1} $\lambda$1304 triplet and the \ion{Si}{2} $\lambda$1308 doublet blended in the $\lambda$1304
bump typically seen in AGN spectra vary from quasar to quasar, ranging from 20\% to 60\% in the \ion{O}{1} fraction.
Such a large flux in the \ion{Si}{2} doublet might be accounted for by the LOC-type integration over the gas physical parameter space or
by an unusually large microturbulent velocity. % corresponding to $v_{\rm turb} >$ 1000 km s$^{-1}$.

\noindent 4.\ The flux ratio between \ion{O}{1} $\lambda$8446 and $\lambda$11287 observed in Sy1s may indicate that the physical condition
within the line-emitting gas is similar to those in quasars.
On the other hand, the ratios observed in NLS1s suggest that radiation-related processes are more efficient over collisional processes than 
in quasars and Sy1s.
It might be a consequence of a stronger continuum radiation incident on the line-emission region, or lower gas density in NLS1s.

\noindent 5.\ In the LOC scenario, the revealed physical parameters of the line-emitting gas should represent the maximum contribution to the 
observed \ion{O}{1} and \ion{Ca}{2} emission in the parameter space.
Such a condition requires the distribution functions of the gas covering fraction and the density of $f(r) \propto r^{-1}$ and 
$g(n_{\rm H}) \propto n_{\rm H}^{-1}$, respectively.


%% Included in this acknowledgments section are examples of the
%% AASTeX hypertext markup commands. Use \url without the optional [HREF]
%% argument when you want to print the url directly in the text. Otherwise,
%% use either \url or \anchor, with the HREF as the first argument and the
%% text to be printed in the second.

\acknowledgments

We are grateful to an anonymous referee for giving a number of useful comments to improve this paper.
G. J. Ferland kindly gave us the useful advices regarding the photoionization code Cloudy.
We thank the staff of UKIRT for technical support and assistance with the observation.
Use of the UKIRT is supported by the National Astronomical Observatory of Japan.
This work has been supported in part by Grants-in-Aid for Scientific Research
(15253002, 17104002) and the Japan-Australia Research Cooperative Program from the Japan Society for
the Promotion of Science.

\begin{table}
\begin{center}
\caption{Characteristics of the Observed Quasars\label{sample}}
\begin{tabular}{cccccccc}
\tableline\tableline
Object & Redshift\tablenotemark{a} & $M_B$\tablenotemark{b} & $A_V$\tablenotemark{c} & $J$ (2MASS)\tablenotemark{d} & 
$H$ (2MASS)\tablenotemark{d} & $J$\tablenotemark{e} & $H$\tablenotemark{e}\\
\tableline
3C 273         &  0.158   & $-$26.9 & 0.007 & 11.8 & 11.0 & 11.5 & \nodata\\
QSO B0850+440  &  0.514   & $-$26.1 & 0.011 & 15.2 & 14.4 & \nodata & 15.0\\
3C 232         &  0.530   & $-$26.7 & 0.005 & 14.9 & 14.4 & \nodata & 14.4\\
QSO J1139$-$1350 &  0.560 & $-$26.3 & 0.013 & 15.5 & 14.9 & \nodata & 15.0\\
PG 1148+549    &  0.978   & $-$27.7 & 0.004 & 14.5 & 14.2 & \nodata & 14.6\\
PG 1718+481    &  1.084   & $-$29.8 & 0.006 & 13.5 & 13.0 & \nodata & 13.0\\
\tableline
\end{tabular}
\tablenotetext{a}{Redshift measured in the UV spectra \citep{evans04}.}
\tablenotetext{b}{The $B$-band absolute magnitude taken from \citet{veron03}.}
\tablenotetext{c}{Galactic extinction \citep{SFD98}; $R_{\rm V}$ = 3.08 was adopted \citep{pei92}.}
\tablenotetext{d}{The $J$- and $H$-band magnitudes measured in the 2MASS observations.}
\tablenotetext{e}{The $J$- and $H$-band magnitudes measured in this work.}
\end{center}
\end{table}

\begin{table}
\begin{center}
\caption{Observing Journal of the Near-IR Spectroscopy\label{irsp}}
\begin{tabular}{cccccc}
\tableline\tableline
Target & Exposure Time (s) & Grism & Air Mass & Rationing Star\\
\tableline
3C 273          & 2400 & $IJ$ & 1.1 & BS 4708 (F8V)\\
QSO B0850+440   & 3840 & $JH$ & 1.5 & BS 3451 (F7V)\\
3C 232          & 7680 & $JH$ & 1.2 & BS 3625 (F9V)\\
QSO J1139$-$1350  & 6720 & $JH$ & 1.3 & BS 4529 (F7V)\\
PG 1148+549     & 2880 & $HK$ & 1.2 & BS 4761 (F7V)\\
PG 1718+481     & 1920 & $HK$ & 1.5 & BS 6467 (F4V)\\
\tableline
\end{tabular}
\end{center}
\end{table}

\begin{table}
\begin{center}
\caption{Observing Journal of the Near-IR Imaging\label{irim}}
\begin{tabular}{cccccc}
\tableline\tableline
Target & Exposure Time (s) & Filter & Air Mass & Standard Star\\
\tableline
3C 273          & 50 & $J$ & 1.1 & FS 132\\
QSO B0850+440   & 300 & $H$ & 1.3 & FS 125\\
3C 232          & 300 & $H$ & 1.6 & FS 127\\
QSO J1139$-$1350  & 300 & $H$ & 1.6 & FS 129\\
PG 1148+549     & 300 & $H$ & 1.2 & FS 121\\
PG 1718+481     & 150 & $H$ & 1.6 & FS 141\\
\tableline
\end{tabular}
\end{center}
\end{table}

\begin{deluxetable}{ccccccc}
\tabletypesize{\scriptsize}
\rotate
\tablecaption{Measured Emission-Line Fluxes\tablenotemark{a}\label{lineflux}}
\tablewidth{0pt}
\tablehead{
\colhead{Emission Lines} & \colhead{3C 273} & \colhead{QSO B0850+440} & \colhead{3C 232} & \colhead{QSO J1139$-$1350} &
\colhead{PG 1148+549} & \colhead{PG 1718+481}
}
\startdata
\cutinhead{UV Lines}
Ly$\beta$ + \ion{O}{6} $\lambda$1030       & 2530 $\pm$ 140  & 160 $\pm$ 24    & \nodata           & 123 $\pm$ 10      & 97.1 $\pm$ 9.9    & 123 $\pm$ 9    \\
%Ly$\alpha$ $\lambda$1216                   & 16000 $\pm$ 200 & 630 $\pm$ 26    & 356 $\pm$ 24      & 980 $\pm$ 10     & 1100 $\pm$ 20      & 1740 $\pm$ 40  \\
%                                                                                                    % NV subtracted.
%\ion{N}{5} $\lambda$1240                   & 3600 $\pm$ 160  & 80.1 $\pm$ 11.9 & 101 $\pm$ 20      & 20.5 $\pm$ 2.4    & 104 $\pm$ 11      & 318 $\pm$ 30   \\
Ly$\alpha$ + \ion{N}{5} $\lambda$1220      & 19600 $\pm$ 200 & 711 $\pm$ 26    & 457 $\pm$ 16      & 1030 $\pm$ 20     & 1200 $\pm$ 20     & 2070 $\pm$ 30\\
\ion{Si}{2} $\lambda$1264                  & 53.5 $\pm$ 11.2 & 5.88 $\pm$ 0.55 & 3.60 $\pm$ 1.66   & $<$7.7           & 14.4 $\pm$ 6.3    & 25.6 $\pm$ 2.3 \\
                                                                                                    % 1 sigma = 2.55
\ion{O}{1} + \ion{Si}{2} $\lambda$1304     & 404 $\pm$ 50    & 24.3 $\pm$ 3.0  & 24.7 $\pm$ 3.0    & 19.2 $\pm$ 2.0    & 37.3 $\pm$ 4.2    & 45.0 $\pm$ 8.2 \\
% This line is total flux of the 1304 feature, not corrected for SiII contributions.
%\ion{O}{1} $\lambda$1304 (a)               & 388 $\pm$ 50    & 22.5 $\pm$ 3.0  & 23.6 $\pm$ 3.0    & 19.2 $\pm$ 2.0    & 33.0 $\pm$ 4.6    & 37.3 $\pm$ 8.2 \\
%     % SiII components subtracted.
%\ion{O}{1} $\lambda$1304 (b)               & 202 $\pm$ 25    & 12.2 $\pm$ 1.5  & 12.4 $\pm$ 1.5    &  9.6 $\pm$ 1.0    & 18.7 $\pm$ 2.1    & 22.5 $\pm$ 4.1 \\
%     % 50% of the 1304 feature.
\ion{C}{2} $\lambda$1335                   & \nodata         & 10.2 $\pm$ 1.3  & 9.94 $\pm$ 1.12   & 7.01 $\pm$ 1.68   & 24.1 $\pm$ 3.2    & 20.3 $\pm$ 3.7 \\
\ion{Si}{4} + \ion{O}{6}] $\lambda$1400    & \nodata         & 70.1 $\pm$ 3.4  & 41.1 $\pm$ 2.36   & 68.6 $\pm$ 7.4    & 101 $\pm$ 6       & 174 $\pm$ 14   \\
\ion{C}{4} $\lambda$1549                   & 7400 $\pm$ 140  & 263 $\pm$ 7     & 165 $\pm$ 6       & 578 $\pm$ 7       & 216 $\pm$ 8       & 196 $\pm$ 31   \\
\ion{C}{3}] $\lambda$1909                  & 1680 $\pm$ 90   & 112 $\pm$ 7     & 128 $\pm$ 5       & 51.9 $\pm$ 3.3    & \nodata           & \nodata        \\
\cutinhead{Near-IR Lines}
\ion{O}{1} $\lambda$8446                   & 239 $\pm$ 26    & 5.07 $\pm$ 1.17 & 0.98 $\pm$ 0.17 & 4.31 $\pm$ 0.77   & 8.10 $\pm$ 0.26   & 9.62 $\pm$ 1.05 \\
%\ion{O}{1} $\lambda$8446                   & 205 $\pm$ 22    & 4.97 $\pm$ 1.15 & 0.981 $\pm$ 0.170 & 3.70 $\pm$ 0.66   & 7.34 $\pm$ 0.24   & 8.52 $\pm$ 0.93 \\
% Not corrected for flux falling in outside the fitting range.
\ion{Ca}{2} $\lambda$8579\tablenotemark{b} & 343 $\pm$ 31    & 8.15 $\pm$ 1.83 & $<$0.39          & 2.08 $\pm$ 0.59   & 9.40 $\pm$ 0.29   & 10.5 $\pm$ 1.7 \\
                                                                                % 1 sigma = 0.130                                           
%\ion{Ca}{2} $\lambda$8579\tablenotemark{b} & 310 $\pm$ 28    & 8.11 $\pm$ 1.82 & $<$ 0.39          & 1.80 $\pm$ 0.51   & 8.78 $\pm$ 0.27   & 4.87 $\pm$ 0.77 \\
%                                                                                % 1 sigma = 0.130                                           
% Not corrected for flux falling in outside the fitting range.
Pa$\epsilon$ +[\ion{S}{3}] $\lambda$9540   & 228 $\pm$ 12    & 4.49 $\pm$ 0.54 & 5.61 $\pm$ 0.20   & 5.60 $\pm$ 0.47   & 6.51 $\pm$ 0.35   & \nodata         \\
Pa$\delta$ + \ion{Fe}{2} $\lambda$10050    & 302 $\pm$ 7     & 5.38 $\pm$ 0.37 & 2.35 $\pm$ 0.22   & 10.5 $\pm$ 0.8    & 9.01 $\pm$ 0.29   & 17.3 $\pm$ 0.48 \\
\ion{Fe}{2} $\lambda$10501                 & \nodata         & 1.95 $\pm$ 0.28 & \nodata           & 1.25 $\pm$ 0.35   & 0.99 $\pm$ 0.13 & \nodata         \\
\ion{He}{1} $\lambda$10830                 & 499 $\pm$ 20    & 14.2 $\pm$ 1.4  & 15.0 $\pm$ 1.3    & 23.2 $\pm$ 3.6    & 23.9 $\pm$ 13.1   & 25.6 $\pm$ 1.9  \\
Pa$\gamma$ $\lambda$10941                  & 402 $\pm$ 20    & 4.52 $\pm$ 1.15 & 9.76 $\pm$ 1.27   & 11.5 $\pm$ 3.6    & 16.5 $\pm$ 13.1   & 35.9 $\pm$ 2.1  \\
\ion{O}{1} $\lambda$11287                  & 59.4 $\pm$ 1.2  & 2.34 $\pm$ 0.20 & 1.29 $\pm$ 0.57   & 0.77 $\pm$ 0.11 & 1.99 $\pm$ 0.10   & 3.41 $\pm$ 0.76 \\
\enddata
%% Text for table notes should follow after the \enddata but before
%% the \end{deluxetable}. Make sure there is at least one \tablenotemark
%% in the table for each \tablenotetext.
\tablenotetext{a}{\rm{T}he fluxes are given in units of 10$^{-15}$ erg s$^{-1}$ cm$^{-2}$.}
\tablenotetext{b}{Total flux of the near-IR \ion{Ca}{2} triplet ($\lambda$8498, $\lambda$8542, and $\lambda$8662).}
\end{deluxetable}

\begin{deluxetable}{ccccc}
\tabletypesize{\scriptsize}
\rotate
\tablecaption{\ion{O}{1} and \ion{Ca}{2} Line Strengths Used to Constrain the Models\label{lineratios}}
\tablewidth{0pt}
\tablehead{
\colhead{Object} & \colhead{EW (\ion{O}{1} $\lambda$8446) (\AA)}
& \colhead{\ion{O}{1} $n$($\lambda$11287)/$n$($\lambda$8446)}
& \colhead{$n$(\ion{Ca}{2})/$n$(\ion{O}{1} $\lambda$8446)\tablenotemark{a}}
& \colhead{References\tablenotemark{b}}}
\startdata
\cutinhead{Quasar}
3C 273         & 19.9   & 0.33 $\pm$ 0.04 & 1.46 $\pm$ 0.21 & This work\\ 
QSO B0850+440  & 37.7   & 0.62 $\pm$ 0.15 & 1.63 $\pm$ 0.53 & This work\\
3C 232\tablenotemark{c} &  2.7 & 1.76 $\pm$ 0.83 & $<$ 0.40       & This work\\
QSO J1139$-$1350 & 25.0   & 0.24 $\pm$ 0.05 & 0.49 $\pm$ 0.16 & This work\\
PG 1148+549    & 51.5   & 0.33 $\pm$ 0.02 & 1.18 $\pm$ 0.05 & This work\\
PG 1718+481    & 15.9   & 0.47 $\pm$ 0.12 & 1.11 $\pm$ 0.22 & This work\\ 
PG 1116+215    & 23.8   & 0.76 $\pm$ 0.27 & \nodata & 1\\
\cutinhead{Sy1}
NGC 863        & \nodata &  0.55 $\pm$ 0.08 & \nodata & 2 \\
\cutinhead{NLS1}
1H 1934$-$063  & \nodata &  0.64 $\pm$ 0.05 & \nodata & 2 \\
Ark 564        & \nodata &  0.82 $\pm$ 0.03 & \nodata & 2 \\
Mrk 335        & \nodata &  0.64 $\pm$ 0.05 & \nodata & 2 \\
Mrk 1044       & \nodata &  0.42 $\pm$ 0.05 & \nodata & 2 \\
Ton S180       & \nodata &  1.08 $\pm$ 0.16 & \nodata & 2 \\
I Zw 1         & \nodata &  0.76 $\pm$ 0.11 & \nodata & 3, 4\\
\enddata
%% Text for table notes should follow after the \enddata but before
%% the \end{deluxetable}. Make sure there is at least one \tablenotemark
%% in the table for each \tablenotetext.
\tablenotetext{a}{The representative wavelength of 8579 \AA\ was used to convert the \ion{Ca}{2} triplet flux to the photon number flux.}
\tablenotetext{b}{References --- (1) \citet{matsuoka05}. (2) \citet{rodriguez02b}. (3) \citet{laor97}. (4) \citet{rudy00}.}
\tablenotetext{c}{This quasar is excluded from the sample to be compared with the model calculations, due to its anomalously weak \ion{O}{1} $\lambda$8446 emission.}
\end{deluxetable}

\begin{table}
\begin{center}
\caption{Input Parameters of the Photoionized Models \label{models}}
\begin{tabular}{ccc}
\tableline\tableline
Model & Continuum Shape & $v_{\rm turb}$ \\
 & ($T_{\rm cut}$ [K], $\alpha_{\rm uv}$, $\alpha_{\rm x}$, $\alpha_{\rm ox}$)\tablenotemark{a} & (km s$^{-1}$) \\
\tableline
1  (standard) & (1.5 $\times$ 10$^5$, $-$0.2, $-$1.8, $-$1.4) & 0    \\
2             & (1.5 $\times$ 10$^5$, $-$0.2, $-$1.8, $-$1.4) & 10   \\
3             & (1.5 $\times$ 10$^5$, $-$0.2, $-$1.8, $-$1.4) & 100  \\
4             & (1.0 $\times$ 10$^6$, $-$0.5, $-$1.0, $-$1.4) & 100  \\
\tableline
\end{tabular}
%% Any table notes must follow the \end{tabular} command.
\end{center}
\tablenotetext{a}{See text for the definition of these parameters.}
\end{table}

\begin{table}
\begin{center}
\caption{Results of the Photoionized Model Calculations and the Compared Observation\label{bfmodels}}
\begin{tabular}{cccccc}
\tableline\tableline
Parameter/Prediction  & Model 1 & Model 2 & Model 3 & Model 4 & Observed\\
\tableline
\cutinhead{Best-Fit Parameter}
$n_{\rm H}$ (cm$^{-3}$) & 10$^{11.5}$ & 10$^{11.5}$ & 10$^{12.0}$ & 10$^{12.0}$ & N/A\\
$U$                     & 10$^{-3.0}$ & 10$^{-2.5}$ & 10$^{-2.5}$ & 10$^{-2.5}$ & N/A\\
\cutinhead{Predicted Line Strengths}
EW (\ion{O}{1} $\lambda$8446)\tablenotemark{a} & 12.0 & 12.1 & 8.72 & 5.08 & $>$ 10\\
\ion{O}{1} $n$($\lambda$11287)/$n$($\lambda$8446)    & 0.56 & 0.58 & 0.52 & 0.53 & 0.46 $\pm$ 0.18\\
\ion{O}{1} $n$($\lambda$1304)/$n$($\lambda$8446)     & 0.45 & 0.48 & 0.40 & 0.41 & \nodata\\
$n$(\ion{Ca}{2})/$n$(\ion{O}{1} $\lambda$8446)       & 2.51 & 1.57 & 0.95 & 1.04 & 0.1 - 5.0\\
\tableline
\end{tabular}
\tablenotetext{a}{The EWs are computed with a covering factor of 0.1 and given in units of angstroms.}
\end{center}
\end{table}

\begin{deluxetable}{ccccccc}
\tabletypesize{\scriptsize}
\rotate
\tablecaption{Observed and Theoretical Values of the \ion{O}{1} Line Ratios \label{otheroi}}
\tablewidth{0pt}
\tablehead{
\colhead{Object} & type & \colhead{$\lambda$7254/$\lambda$8446}
& \colhead{$\lambda$7774/$\lambda$8446} & \colhead{$\lambda$7990/$\lambda$8446}
& \colhead{$\lambda$13165/$\lambda$11287} & \colhead{References\tablenotemark{a}}}
\startdata
\cutinhead{Observed}
NGC 4151 & Sy1 & $\le$ 0.08 & $\le$ 0.02 & $\le$ 0.04 & \nodata & 1\\
NGC 5548 & Sy1 & $\le$ 0.05 & $\le$ 0.06 & $\le$ 0.02 & \nodata & 1\\
Mrk 42   & Sy1 & $\le$ 0.03 & $\le$ 0.08 & $\le$ 0.03 & \nodata & 1\\
Mrk 335  & Sy1 & $\le$ 0.01 & $\le$ 0.04 & $\le$ 0.02 & \nodata & 1\\
Akn 564  & Sy1 & $\le$ 0.04 & $\le$ 0.03 & $\le$ 0.02 & \nodata & 1\\
1H 1934$-$063 & NLS1 & $<$ 0.03 & 0.07 & 0.04 & $<$ 0.02 & 2\\
Ark 564     & NLS1 & \nodata & \nodata & \nodata & $<$ 0.07 & 2\\
Mrk 335     & NLS1 & \nodata & \nodata & \nodata & $<$ 0.05& 2\\
Mrk 1044    & NLS1 & \nodata & \nodata & \nodata & $<$ 0.14 & 2\\
Ton S180    & NLS1 & \nodata & $<$ 0.08 & $<$ 0.09 & \nodata & 2\\
\cutinhead{Theoretical Predictions}
Ly$\beta$ fluorescence &     & 0.0   & 0.0     & 0.0   & 0.0       & --\\
Collisional excitation &     & \nodata & 0.3\tablenotemark{b} & \nodata & \nodata & 1\\
Continuum fluorescence &     & 0.025 & \nodata & 0.052 & $\sim$ 1.0 & 1\\
Recombination          &     & \nodata & 7.2\tablenotemark{c} & \nodata & \nodata & This work\\
\enddata
%% Text for table notes should follow after the \enddata but before
%% the \end{deluxetable}. Make sure there is at least one \tablenotemark
%% in the table for each \tablenotetext.
\tablenotetext{a}{References --- (1) \citet{grandi80}. (2) \citet{rodriguez02b}.}
\tablenotetext{b}{Only the collisional cross sections from the ground term are considered (see text).}
\tablenotetext{c}{Appropriate in the low-density limit (e.g., in the normal Galactic gaseous nebulae).}
\end{deluxetable}

\begin{table}
\begin{center}
\caption{Constitution of the $\lambda$1304 Bump\tablenotemark{a} \label{oi1304}}
\begin{tabular}{ccc}
\tableline\tableline
Object & \ion{O}{1} \tablenotemark{b} & \ion{Si}{2}\\
\tableline
3C 273          &  264 (0.65) -- 396 (0.98)  &    8 -- 140 \\
QSO B0850+440   & 15.3 (0.63) -- 18.5 (0.76) &  5.8 -- 9.0 \\
QSO J1139$-$1350  &  1.7 (0.09) -- 4.1  (0.21) & 15.1 -- 17.5 \\
PG 1148+549     &  7.8 (0.21) -- 12.6 (0.34) & 24.7 -- 29.5 \\
PG 1718+481     & 19.8 (0.44) -- 25.4 (0.56) & 19.6 -- 25.2 \\
\tableline
\end{tabular}
\tablenotetext{a}{\rm{F}luxes are given in units of 10$^{-15}$ ergs s$^{-1}$ cm$^{-2}$, as in Table \ref{lineflux}.}
\tablenotetext{b}{Values in parentheses represent the fractions in the bump.}
\end{center}
\end{table}

\begin{table}
\begin{center}
\caption{Measured FWHMs of the \ion{C}{4} $\lambda$1549 and \ion{O}{1} $\lambda$11287 Lines \label{fwhm}}
\begin{tabular}{crr}
\tableline\tableline
Object & \ion{C}{4} $\lambda$1549 & \ion{O}{1} $\lambda$11287\\
\tableline
3C 273          & 3940 & 2160 \\ % HeI 2660
QSO B0850+440   & 5540 & 1960 \\ % HeI 4070
3C 232          & 7430 & 2410 \\ % HeI 1920
QSO J1139$-$1350  & 3790 & 1210 \\ % HeI 1850
PG 1148+549     & 5860 & 1830 \\ % HeI 6280
PG 1718+481     & 3720 & 1840 \\ % HeI 1950
\tableline
\end{tabular}
\end{center}
Note --- The FWHMs are given in units of kilometers per second.
\end{table}

\begin{table}
\begin{center}
\caption{Equivalent Widths of the Typical UV and Optical Emission Lines\label{otherlines}}
\begin{tabular}{ccc}
\tableline\tableline
Line & Model\tablenotemark{a} & Observed\tablenotemark{b}\\
\tableline
Ly$\alpha$ $\lambda$1216 + \ion{N}{5} $\lambda$1240  & 40.8   & 52\\
\ion{Si}{4}+\ion{O}{6}] $\lambda$1400                & 0.49   & 10\\
\ion{C}{4} $\lambda$1549                             & 1.35   & 37\\
\ion{C}{3}] $\lambda$1909                            & 0.23   & 22\\
\ion{Fe}{2} $\lambda\lambda$2240 -- 2650             & 16.7   & 49\\
\ion{Mg}{2} $\lambda$2800                            & 65.0   & 50\\
H$\beta$ $\lambda$4863                               & 31.0   & 58\\
$[$\ion{O}{3}] $\lambda$5007                         & 0.0020 & 15\\
\tableline
\end{tabular}
\tablenotetext{}{Note --- Equivalent widths are given in units of angstroms.}
\tablenotetext{a}{The standard model with the best-fit ($n_{\rm H}$, $U$) parameters to the \ion{O}{1} and \ion{Ca}{2} observations and a covering factor of 0.1.}
\tablenotetext{b}{The EWs of Emission lines in the LBQS composite spectrum were taken from \citet{francis91}, except for that of the \ion{Fe}{2},
  which was taken from \citet{baldwin04}.}
\end{center}
\end{table}

\clearpage

\begin{figure}
\epsscale{0.8}
\plotone{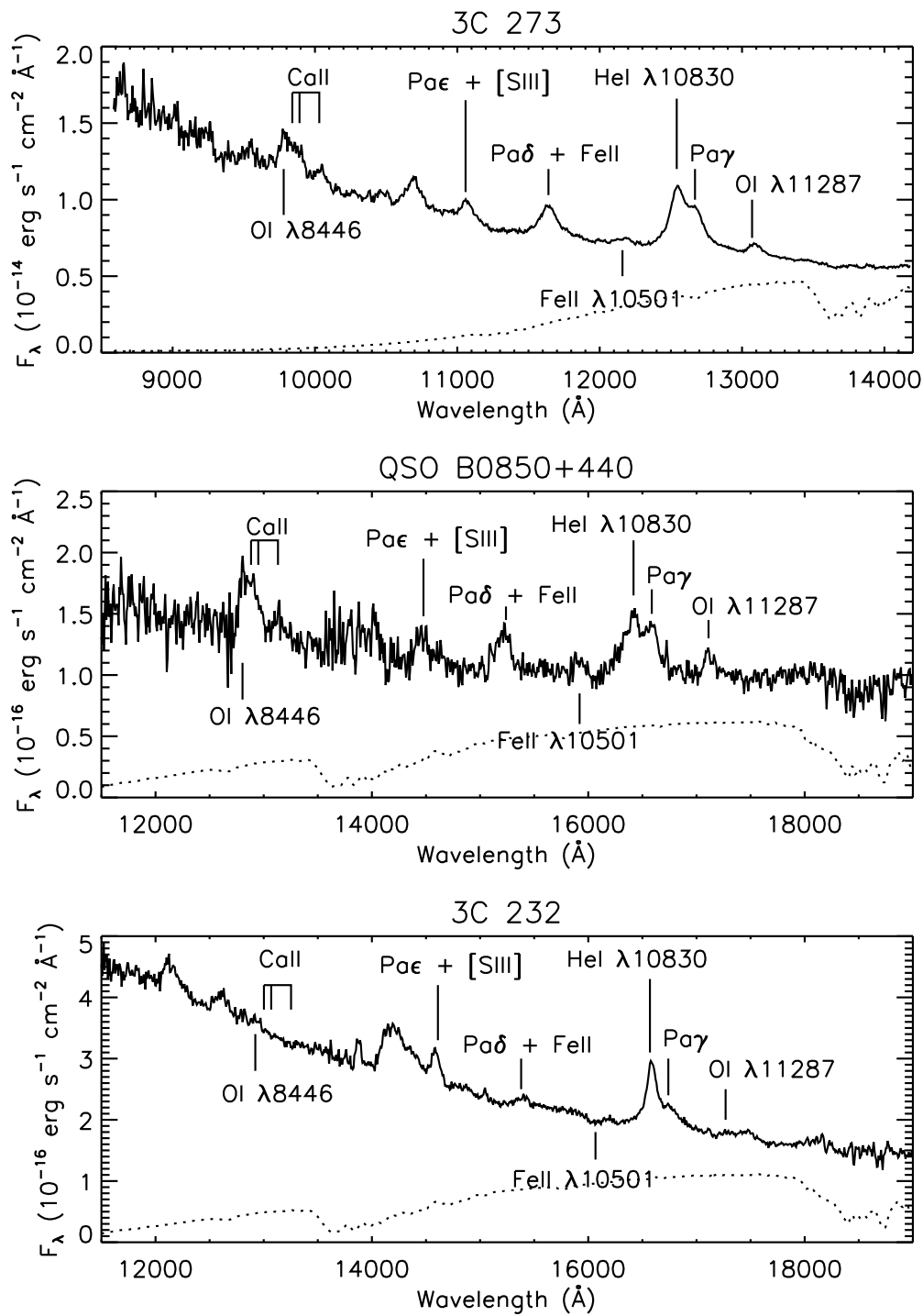}
\caption{Reduced near-IR spectra ({\it solid lines}) along with the sensitivity curves 
  (with arbitrary amounts of scaling; {\it dotted lines}) for 3C 273, QSO B0850+440, and 3C 232 in the observed frame. 
  Positions of the identified emission lines are marked with vertical lines.\label{plot1}}
\end{figure}

\begin{figure}
\epsscale{0.8}
\plotone{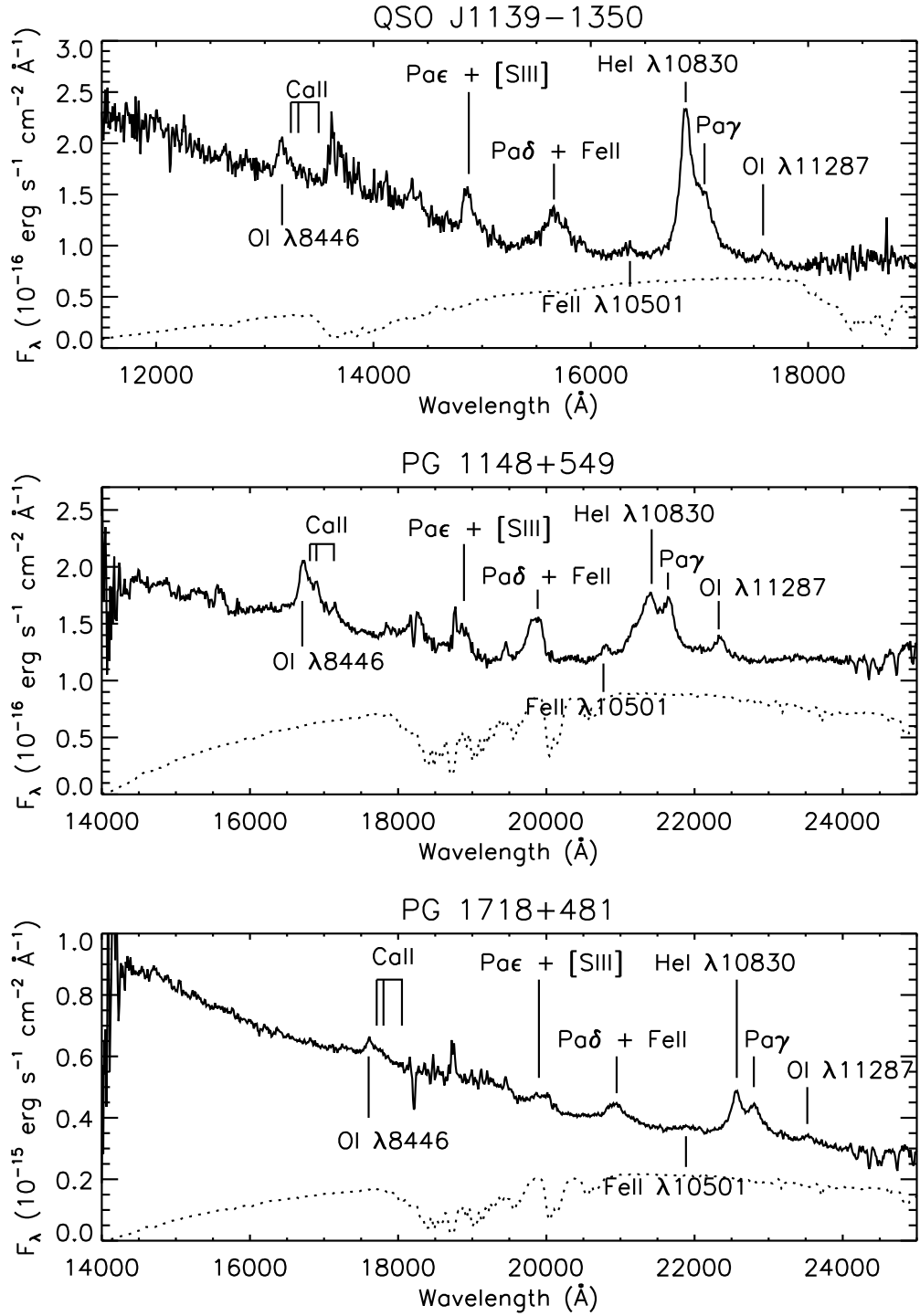}
\caption{Same as Fig. \ref{plot1} but for QSO J1139$-$1350, PG 1148+549, and PG 1718+481.\label{plot2}}
\end{figure}

\begin{figure}
\epsscale{1.0}
\plotone{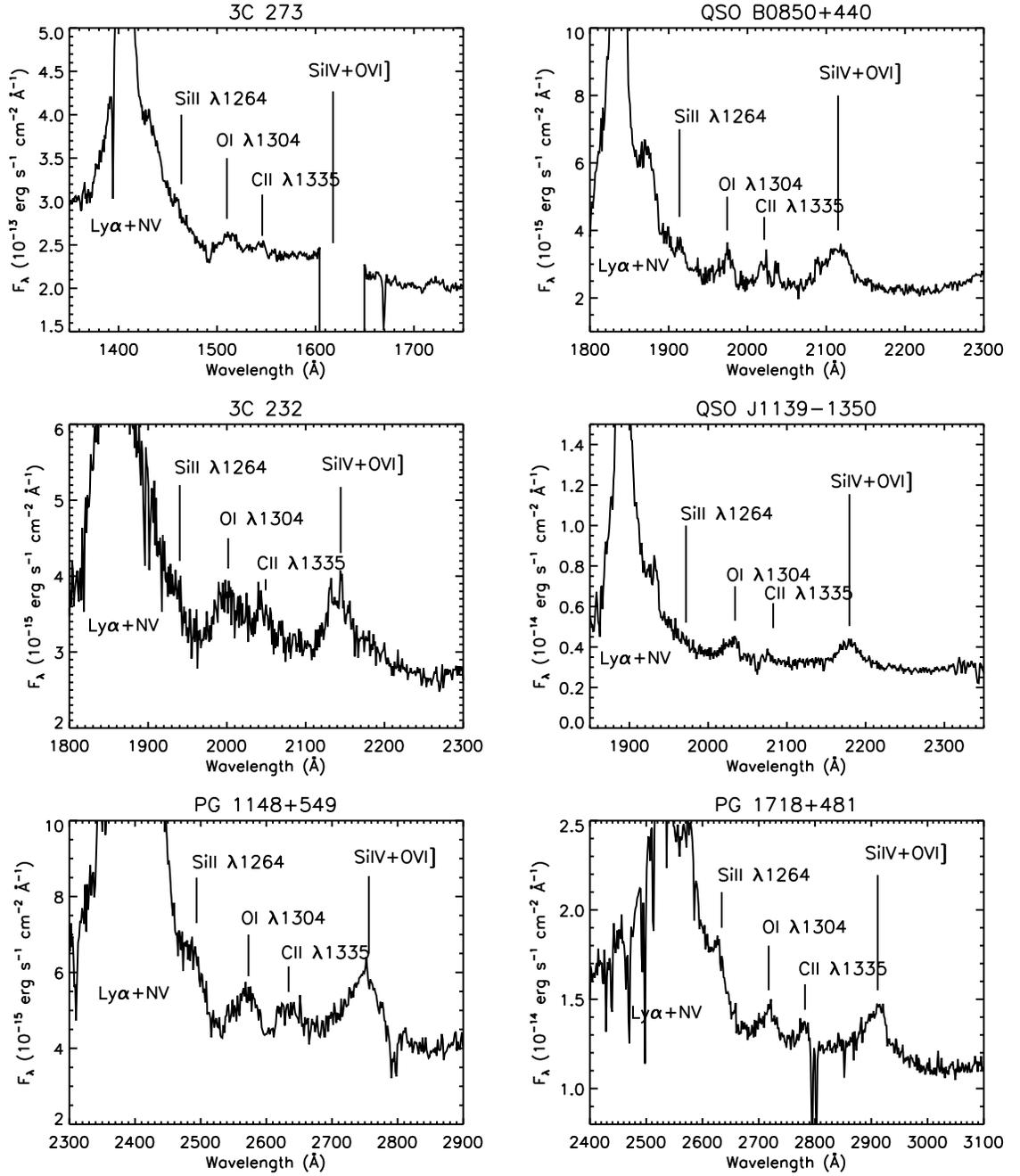}
\caption{UV spectra around \ion{O}{1} $\lambda$1304 for 3C 273, QSO B0850+440, 3C 232, QSO J1139$-$1350, PG 1148+549, 
  and PG 1718+481 in the observed frame. 
  Positions of the identified emission lines are marked with vertical lines.\label{plot_uv}}
\end{figure}

\begin{figure}
\epsscale{0.4}
\plotone{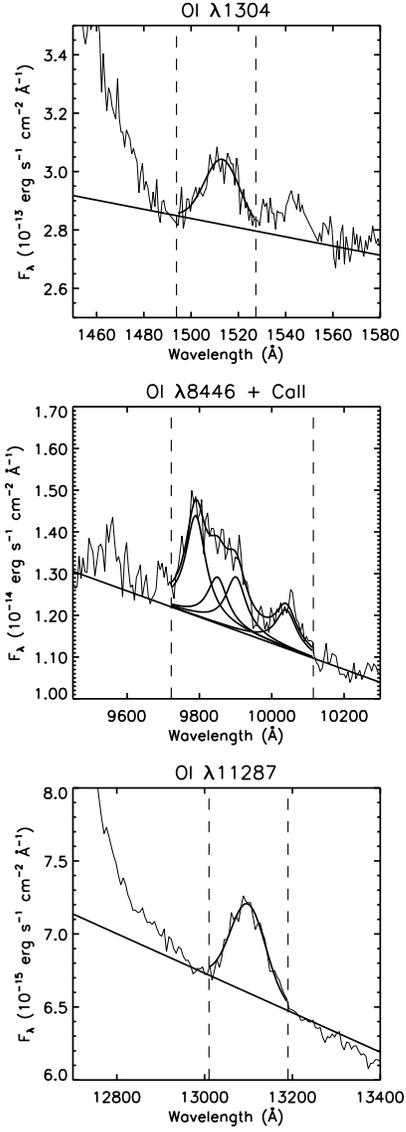}
\caption{Fitting results for \ion{O}{1} $\lambda$1304, \ion{O}{1} $\lambda$8446 + the \ion{Ca}{2} triplet
($\lambda$8498, $\lambda$8542, and $\lambda$8662), and \ion{O}{1} $\lambda$11287 in 3C 273. 
  Observed spectra are shown by thin solid lines while the determined continuum and the fitted Voigt profiles are shown by thick
  solid lines.
  The sum of these fitted components are also shown by a thick solid line for \ion{O}{1} $\lambda$8446 + the \ion{Ca}{2} triplet.
  The wavelength ranges used in the fitting are indicated by vertical dashed lines. \label{fit}}
\end{figure}

\begin{figure}
\epsscale{0.8}
\plotone{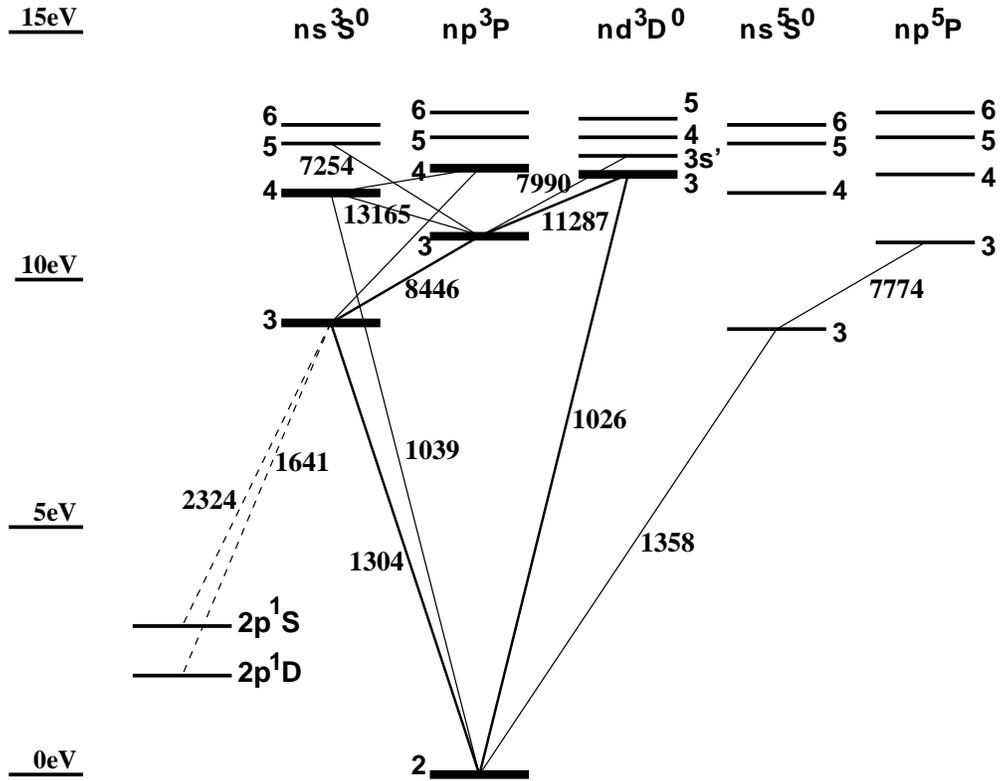}
\caption{Partial Grotrian diagram of an \ion{O}{1} atom. Solid lines represent permitted transitions,
  while dashed lines represent semi-forbidden transitions. 
  Transition paths of the Ly$\beta$-pumped electrons (corresponding to $\lambda$1026, $\lambda$11287, $\lambda$8446,
  and $\lambda$1304) are shown by thick solid lines.
  The six energy levels considered in the model calculations are shown by thick horizontal lines.
\label{oienergy}}
\end{figure}

\begin{figure}
\epsscale{0.8}
\plotone{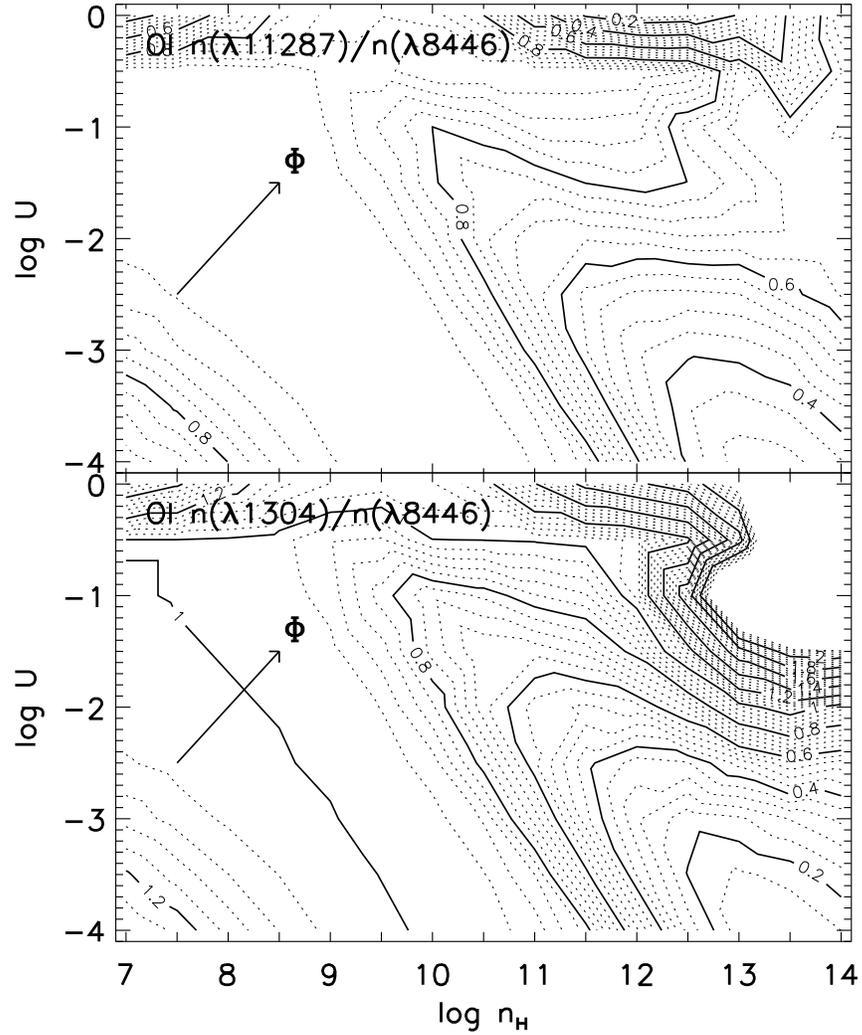}
\caption{\ion{O}{1} photon flux ratios $n$($\lambda$11287)/$n$($\lambda$8446) ({\it top}) and
  $n$($\lambda$1304)/$n$($\lambda$8446) ({\it bottom}) as a function of the gas density $n_{\rm H}$ and 
  the ionization parameter $U$, calculated in the standard model. 
  The contours are linearly spaced. 
  Arrows indicate the direction toward which the incident-ionizing continuum flux $\Phi $ increases.\label{oicaii1}}
\end{figure}

\begin{figure}
\epsscale{1.0}
\plotone{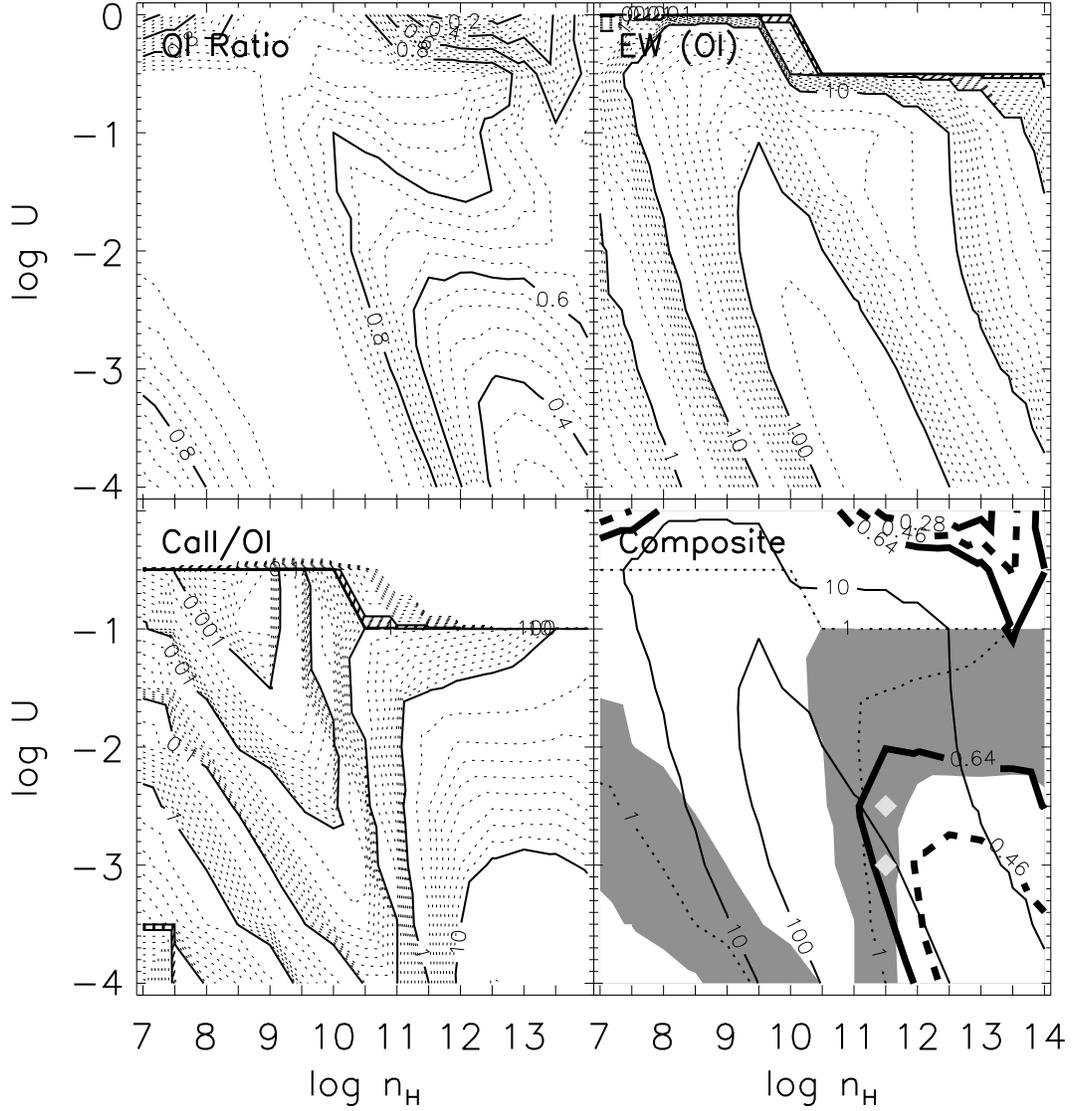}
\caption{\ion{O}{1} $n$($\lambda$11287)/$n$($\lambda$8446) ({\it top left}), EW (\ion{O}{1} $\lambda$8446) in units of angstroms ({\it top right}),
  $n$(\ion{Ca}{2})/$n$(\ion{O}{1} $\lambda$8446) ({\it bottom left}), and their composite plot ({\it bottom right}) as a function of 
  the gas density $n_{\rm H}$ and the ionization parameter $U$, calculated in the standard model.
  In the bottom right panel, the three constraints from the observed \ion{O}{1} $n$($\lambda$11287)/$n$($\lambda$8446), EW (\ion{O}{1} $\lambda$8446),
  and $n$(\ion{Ca}{2})/$n$(\ion{O}{1} $\lambda$8446) values are shown by thick solid lines, thin solid lines, and the shaded area, 
  respectively.
  The models consistent with all the constraints are marked with diamonds.
  Dashed lines and dotted lines represent mean values of the observed \ion{O}{1} $n$($\lambda$11287)/$n$($\lambda$8446)
  and $n$(\ion{Ca}{2})/$n$(\ion{O}{1} $\lambda$8446), respectively.
  Note that the contours in the top left panel are linearly spaced, while they are spaced logarithmically in the top
  right and the bottom left panels.  \label{4chartA}}
\end{figure}

\begin{figure}
\epsscale{0.8}
\plotone{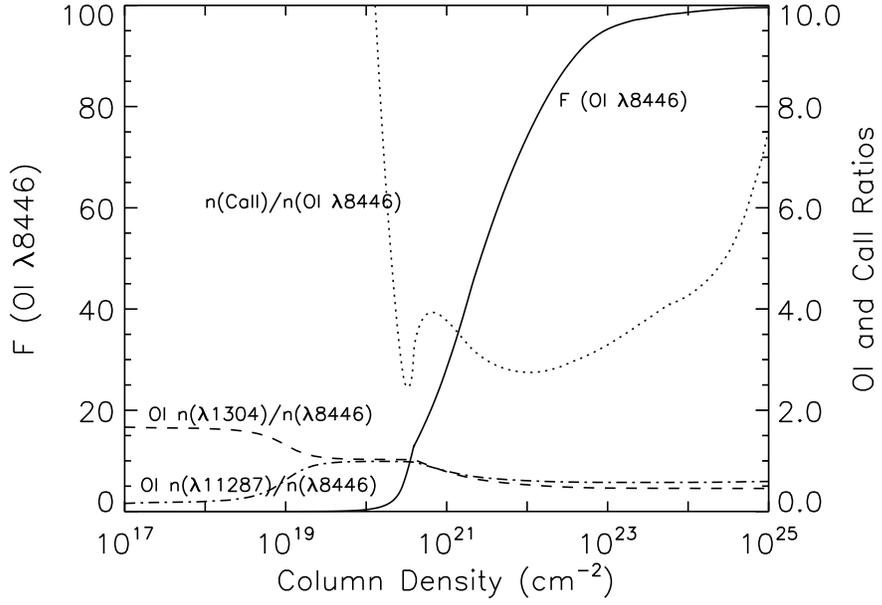}
\caption{Cumulative flux of \ion{O}{1} $\lambda$8446 (with an arbitrary scaling; {\it solid line}) and ratios of \ion{O}{1} 
  $n$($\lambda$1304)/$n$($\lambda$8446), \ion{O}{1} $n$($\lambda$11287)/$n$($\lambda$8446), and
  $n$(\ion{Ca}{2})/$n$(\ion{O}{1} $\lambda$8446) ({\it dashed line, dot-dashed line, and dotted line}, respectively),
  plotted against the gas column density measured from the illuminated surface. 
  The model parameters other than the column density $N_{\rm H}$ used in the calculation are identical to
  those of the standard model with ($n_{\rm H}$, $U$) = (10$^{11.5}$ cm$^{-3}$, 10$^{-3.0}$). \label{cloud_str}}
\end{figure}

\begin{figure}
\epsscale{0.8}
\plotone{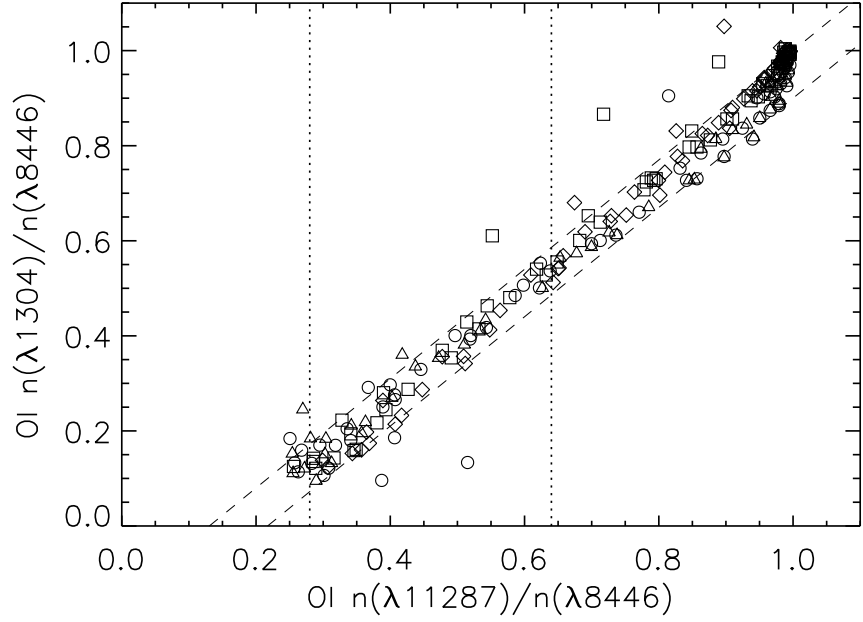}
\caption{\ion{O}{1} photon flux ratios $n$($\lambda$11287)/$n$($\lambda$8446) vs.
  $n$($\lambda$1304)/$n$($\lambda$8446) calculated in models 1 -- 4 with the ($n_{\rm H}$, $U$)
  parameters around the best-fit values, with which the \ion{O}{1} line formation is dominated by Ly$\beta$ fluorescence and collisional
  excitation (see text).
  Diamonds represent model 1 with different ($n_{\rm H}$, $U$) parameter values, while squares represent model 2, circles represent model 3, 
  and triangles represent model 4.
  Dotted lines show a range of the observed values, $n$($\lambda$11287)/$n$($\lambda$8446) = 0.46 $\pm$ 0.18.
  We regard the $n$($\lambda$1304)/$n$($\lambda$8446) values enclosed by two dashed lines, drawn to include most of the points in the observed
  $n$($\lambda$11287)/$n$($\lambda$8446) range,
  as allowed values given the $n$($\lambda$11287)/$n$($\lambda$8446) ratio. \label{oivsoi}}
\end{figure}

\begin{figure}
\epsscale{0.5}
\plotone{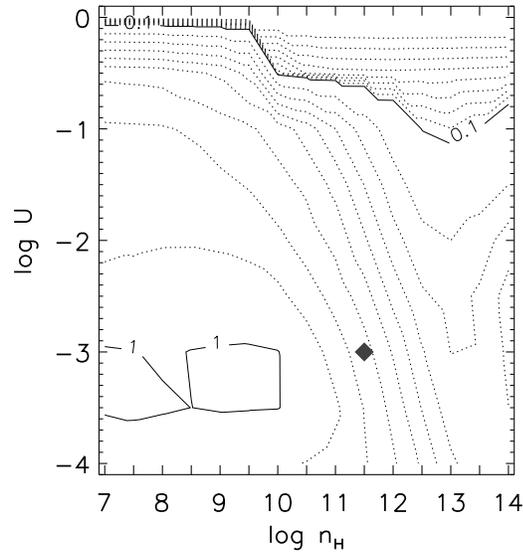}
\caption{EWs of the \ion{Si}{2} $\lambda$1308 doublet in units of angstroms, as a function of 
  the gas density $n_{\rm H}$ and the ionization parameter $U$ calculated in the standard model.
  A diamond represents the best-fit parameters to the \ion{O}{1} and \ion{Ca}{2} observations. \label{ewsi}}
\end{figure}

\begin{figure}
\epsscale{0.8}
\plotone{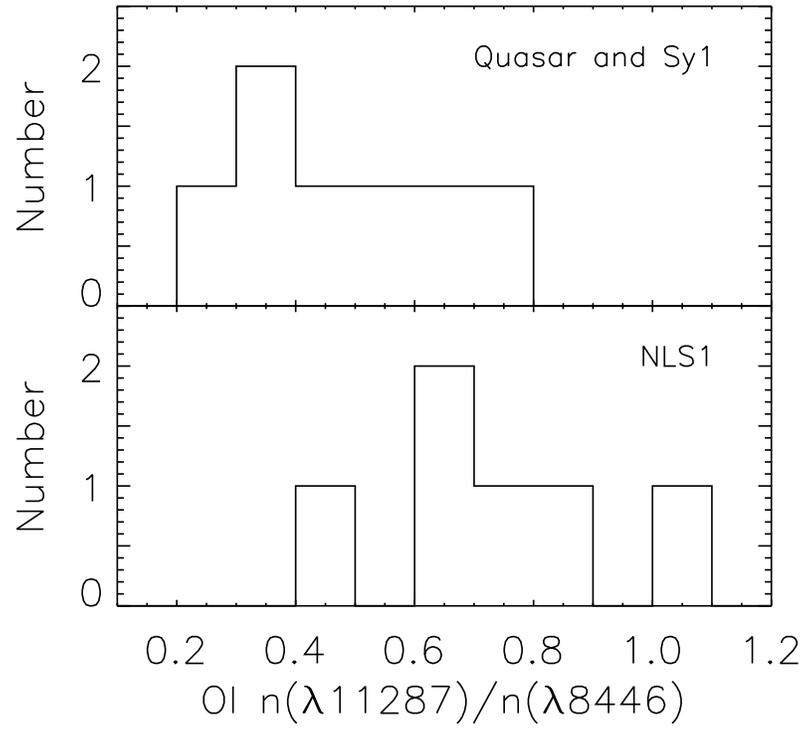}
\caption{Number distributions of the \ion{O}{1} $n$($\lambda$11287)/$n$($\lambda$8446) ratios observed in AGNs.
  The top panel shows the distribution for quasars and Sy1, while the bottom panel shows the same for NLS1s.
  \label{otheragn}}
\end{figure}

\begin{figure}
\epsscale{1.0}
\plotone{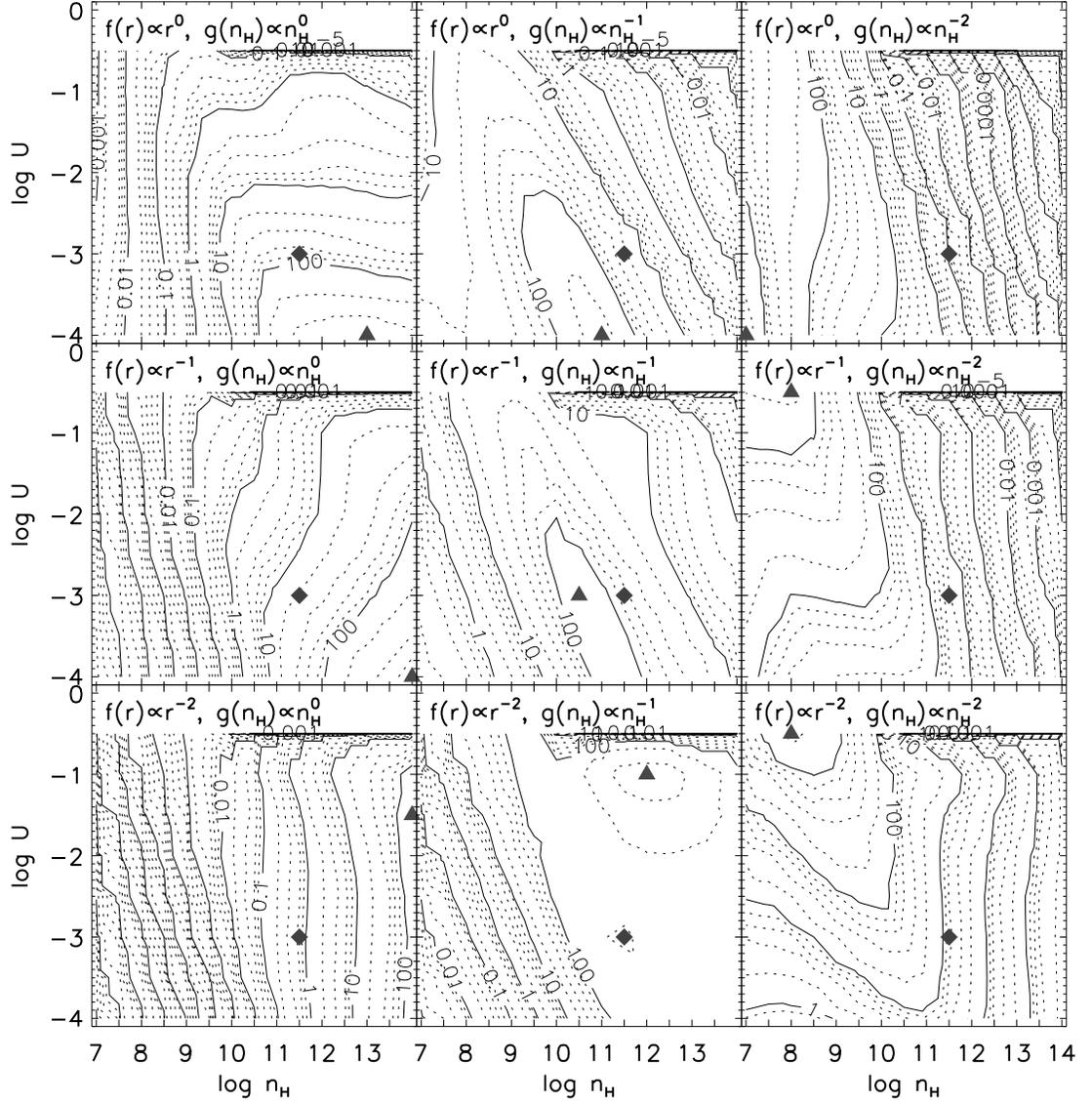}
\caption{Distributions of the DF (see text for its definition) values for \ion{O}{1} $\lambda$8446 as a function of the gas 
  density $n_{\rm H}$ and the ionization parameter $U$ calculated in the standard model.
  Assumed distribution functions of the gas covering fraction and density are indicated at the top of each panel.
  Diamonds represent the best-fit parameters to the \ion{O}{1} and \ion{Ca}{2} observation, while triangles represent the parameters predicting 
  the DF maxima. \label{loc}}
\end{figure}

\begin{figure}
\epsscale{0.5}
\plotone{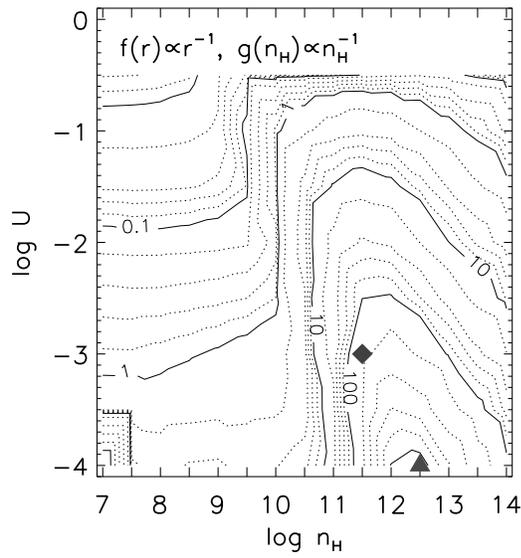}
\caption{Distribution of the DF (see text for its definition) values for \ion{Ca}{2} as a function of the gas density
  $n_{\rm H}$ and the ionization parameter $U$ calculated in the standard model.
  Assumed distribution functions of the gas covering fraction and density are $f(r) \propto r^{-1}$ and 
  $g(n_{\rm H}) \propto n_{\rm H}^{-1}$.
  A diamond represents the best-fit parameters to the \ion{O}{1} and \ion{Ca}{2} observation, while a triangle represents the parameters predicting 
  the DF maximum. \label{loc_ca}}
\end{figure}


\begin{thebibliography}{}

\bibitem[Baldwin et al.(1996)]{baldwin96} Baldwin, J. A., Ferland, G. J., Korista, K. T., Carswell, R. F., Hamann, F., Phillips, M. M., Verner, D.,
  Wilkes, B. J., \& Williams, R. E. 1996, \apj, 461, 664
\bibitem[Baldwin et al.(2004)]{baldwin04} Baldwin, J. A., Ferland, G. J., Korista, K. T., Hamann, F., \& LaCluyz\'{e}, A. 2004, \apj, 615, 610
\bibitem[Baldwin et al.(1995)]{baldwin95} Baldwin, J. A., Ferland, G. J., Korista, K. T., \& Verner, D. 1995, \apj, 455, L119
\bibitem[Binette et al.(1989)]{binette89} Binette, L., Prieto, A., Szuszkiewicz, E., \& Zheng, W. 1989, \apj, 343, 135
\bibitem[Bottorff \& Ferland(2000)]{bf00} Bottorff, M., \& Ferland, G. J. 2000, \mnras, 316, 103
\bibitem[Bottorff et al.(2000)]{bottorff00} Bottorff, M., Ferland, G., Baldwin, J., \& Korista, K. 2000, \apj, 542, 644
\bibitem[Cheng et al.(1991)]{cheng91} Cheng, F. H., Gaskell, C. M., \& Koratkar, A. P. 1991, \apj, 370, 487
\bibitem[Collin \& Joly(2000)]{collin00} Collin, S., \& Joly, M. 2000, \nar, 44, 531
\bibitem[Constantin et al.(2002)]{constantin02} Constantin, A., Shields, J. C., Hamann, F., Foltz, C. B., \& Chaffee, F. H. 2002, \apj, 565, 50
\bibitem[Davidson \& Netzer(1979)]{davidson79} Davidson, K., \& Netzer, H. 1979, Rev. Mod. Phys., 51, 715
\bibitem[Dietrich \& Kollatschny(1995)]{dietrich95} Dietrich, M., \& Kollatschny, W. 1995, \aap, 303, 405
\bibitem[Dietrich et al.(2002)]{dietrich02} Dietrich, M., Appenzeller, I., Vestergaard, M., \& Wagner, S. J. 2002, \apj, 564, 581
\bibitem[Dietrich et al.(2003)]{dietrich03} Dietrich, M., Hamann, F., Appenzeller, I., \& Vestergaard, M. 2003, \apj, 596, 817
\bibitem[De Zotti \& Gaskell(1985)]{dezotti85} De Zotti, G., \& Gaskell, C. M. 1985, \aap, 147, 1
\bibitem[Elston et al.(1994)]{elston94} Elston, R., Thompson, K. L., \& Hill, G. J. 1994, \nat, 367, 250
\bibitem[Evans \& Koratkar(2004)]{evans04} Evans, I. N., \& Koratkar, A. P. 2004, \apjs, 150, 73
\bibitem[Ferland(2004)]{hazy2} Ferland, G. J. 2004, Hazy 2, A brief introduction to Cloudy 96 -- computational methods, 288
\bibitem[Ferland et al.(1998)]{ferland98} Ferland, G. J., Korista, K. T., Verner, D. A., Ferguson, J. W., Kingdon, J. B., \& Verner, E., M. 1998, \pasp, 110, 761
\bibitem[Ferland \& Persson(1989)]{ferland89} Ferland, G. J., \& Persson, S. E. 1989, \apj, 347, 656
\bibitem[Foltz et al.(1987)]{foltz87} Foltz, C. B., Chaffee, F. H., Hewett, P. C., MacAlpine, G. M., Turnshek, D. A., Weymann, R. J., 
  \& Anderson, S. F. 1987, \aj, 94, 1423
\bibitem[Francis et al.(1991)]{francis91} Francis, P. J., Hewett, P. C., Foltz, C. B., Chaffee, F. H., Weymann, R. J., \& Morris, S. L. 1991, \apj, 373, 465
\bibitem[Freudling et al.(2003)]{freudling03} Freudling, W., Corbin, M. R., \& Korista, K. T. 2003, \apj, 587, L67
\bibitem[Glikman et al.(2006)]{glikman06} Glikman, E., Helfand, D. J., \& White, R. L. 2006, \apj, 640, 579
\bibitem[Grandi(1975a)]{grandi75a} Grandi, S. A. 1975a, \apj, 196, 465
\bibitem[Grandi(1975b)]{grandi75b} Grandi, S. A. 1975b, \apj, 199, L43
\bibitem[Grandi(1980)]{grandi80} Grandi, S. A. 1980, \apj, 238, 10
\bibitem[Grandi(1983)]{grandi83} Grandi, S. A. 1983, \apj, 268, 591
\bibitem[Hamann \& Ferland(1993)]{hamann93} Hamann, F., \& Ferland, G. 1993, \apj, 418, 11
\bibitem[Hamann \& Ferland(1999)]{hamann99} Hamann, F., \& Ferland, G. 1999, \araa, 37,487
\bibitem[Hawarden et al.(2001)]{hawarden01}  Hawarden, T. G., Leggett, S. K., Letawsky, M. B., Ballantyne, D. R., \& Casali, M. M. 2001, \mnras, 325, 563
\bibitem[Hopkins et al.(2004)]{hopkins04} Hopkins, P. F., et al. 2004, \aj, 128, 1112
\bibitem[Iwamuro et al.(2004)]{iwamuro04} Iwamuro, F., Kimura, M., Eto, S., Maihara, T., Motohara, K., Yoshii, Y., \& Doi, M. 2004, \apj, 614, 69
\bibitem[Iwamuro et al.(2002)]{iwamuro02} Iwamuro, F., Motohara, K., Maihara, T., Kimura, M., Yoshii, Y., \& Doi, M. 2002, \apj, 565, 63
\bibitem[Kawara et al.(1996)]{kawara96} Kawara, K., Murayama, T., Taniguchi, Y., \& Arimoto, N. 1996, \apj, 470, L85
\bibitem[K\"{o}nigl \& Kartje(1994)]{konigl94} K\"{o}nigl, A., \& Kartje, J. F. 1994, \apj, 434, 446
\bibitem[Korista et al.(1997)]{korista97} Korista, K., Baldwin, J., Ferland, G., \& Verner, D. 1997, \apjs, 108, 401
\bibitem[Krolik \& Kallman(1988)]{krolik88} Krolik, J. H., \& Kallman, T. R. 1988, \apj, 324, 714
\bibitem[Kwan \& Krolik(1981)]{kk81} Kwan, J., \& Krolik, J. H. 1981, \apj, 250, 478
\bibitem[Laor et al.(1997a)]{laor97a} Laor, A., Fiore, F., Elvis, M., Wilkes, B. J., \& McDowell, J. C. 1997a, \apj, 477, 93
\bibitem[Laor et al.(1997b)]{laor97} Laor, A., Jannuzi, B. T., Green, R. F., \& Boroson, T. A. 1997b, \apj, 489, 656
\bibitem[Legett et al.(2006)]{legett06}  Leggett, S. K., Currie, M. J., Varricatt, W. P., Hawarden, T. G., Adamson, A. J., Buckle, J., Carroll, T., Davies, J. K., 
  Davis, C. J., Kerr, T. H., Kuhn, O. P., Seigar, M. S., \& Wold, T. 2006, \mnras, 373, 781
\bibitem[Maiolino et al.(2003)]{maiolino03} Maiolino, R., Juarez, Y., Mujica, R., Nagar, N. M., \& Oliva, E. 2003, \apj, 596, L155
\bibitem[Matsuoka et al.(2005)]{matsuoka05} Matsuoka, Y., Oyabu, S., Tsuzuki, Y., Kawara, K., \& Yoshii, Y. 2005, \pasj, 57, 563
\bibitem[Mineshige et al.(2000)]{mineshige00} Mineshige, S., Kawaguchi, T., Takeuchi, M., \& Hayashida, K. 2000, \pasj, 52, 499
\bibitem[Morris et al.(1991)]{morris91} Morris, S. L., Weymann, R. J., Anderson, S. F., Hewett, P. C., Francis, P. J., Foltz, C. B., Chaffee, F. H.,
  \& MacAlpine, G. M. 1991, \aj, 102, 1627
\bibitem[Netzer \& Laor(1993)]{netzer93} Netzer, H., \& Laor, A. 1993, \apj, 404, L51
\bibitem[Netzer \& Wills(1983)]{netzer83} Netzer, H., \& Wills, B. J. 1983, \apj, 275, 445
\bibitem[O'Brien et al.(1995)]{obrien95} O'Brien, P. T., Goad, M. R., \& Gondhalekar, P. M. 1995, \mnras, 275, 1125
\bibitem[Pei(1992)]{pei92} Pei, Y. C. 1992, \apj, 395, 130
\bibitem[P\'{e}quignot et al.(1991)]{pequignot91} P\'{e}quignot, D., Petitjean, P., \& Boisson, C. 1991, \aap, 251, 680
\bibitem[Persson(1988)]{persson88} Persson, S. E. 1988, \apj, 330, 751
\bibitem[Peterson \& Wandel(1999)]{peterson99} Peterson, B. M., \& Wandel, A. 1999, \apj, 521, L95
\bibitem[Peterson \& Wandel(2000)]{peterson00} Peterson, B. M., \& Wandel, A. 2000, \apj, 540, L13
\bibitem[Peterson et al.(2002)]{peterson02} Peterson, B. M., et al. 2002, \apj, 581, 197
\bibitem[Pounds et al.(1995)]{pounds95} Pounds, K. A., Done, C., \& Osborne 1995, \mnras, 277, L5
\bibitem[Ramsay Howatt et al.(2004)]{ramsay04} Ramsay Howatt, S. K., et al. 2004, \procspie, 5492, 1160
\bibitem[Rees(1987)]{rees87} Rees, M. J. 1987, \mnras, 228, 47
\bibitem[Riffel et al.(2006)]{riffel06} Riffel, R., Rodr\'{i}guez-Ardila, A., \& Pastoriza, M. G. 2006, \aap, 457, 61
\bibitem[Rodr\'{\i}guez-Ardila et al.(2002a)]{rodriguez02a} Rodr\'{i}guez-Ardila, A., Viegas, S. M., Pastoriza, M. G., \& Prato, L. 2002a, \apj, 565, 140
\bibitem[Rodr\'{\i}guez-Ardila et al.(2002b)]{rodriguez02b} Rodr\'{i}guez-Ardila, A., Viegas, S. M., Pastoriza, M. G., Prato, L., \& Donzelli, C. J. 2002b, \apj, 572, 94
\bibitem[Rudy et al.(1991)]{rudy91} Rudy, R. J., Erwin, P., Rossano, G. S., \& Puetter, R. C. 1991, \apj, 383, 344
\bibitem[Rudy et al.(2000)]{rudy00} Rudy, R. J., Mazuk, S., Puetter, R. C., \& Hamann, F. 2000, \apj, 539, 166
\bibitem[Rudy et al.(1989)]{rudy89} Rudy, R. J., Rossano, G. S., \& Puetter, R. C. 1989, \apj, 342, 235
\bibitem[Sanders et al.(1989)]{sanders89} Sanders, D. B., Phinney, E. S., Neugebauer, G., Soifer, B. T., \& Matthews, K. 1989, \apj, 347, 29
\bibitem[Schlegel et al.(1998)]{SFD98} Schlegel, D. J., Finkbeiner, D. P., \& Davis, M. 1998, \apj, 500, 525
\bibitem[Sigut \& Pradhan(1998)]{sigut98} Sigut, T. A. A., \& Pradhan, A. K. 1998, \apj, 499, L139
\bibitem[Sigut \& Pradhan(2003)]{sigut03} Sigut, T. A. A., \& Pradhan, A. K. 2003, \apjs, 145, 15
\bibitem[Tsuzuki et al.(2006)]{tsuzuki06} Tsuzuki, Y., Kawara, K., Yoshii, Y., Oyabu, S., Tanab\'{e}, T., \& Matsuoka, Y. 2006, \apj, 650, 57
\bibitem[Tsuzuki et al.(2007)]{tsuzuki06b} Tsuzuki, Y., et al. 2007, in preparation
\bibitem[Verner et al.(1996)]{verner96} Verner, D. A., Verner, E. M., \& Ferland, G. J. 1996, Atomic Data Nucl. Data Tables, 64, 1
\bibitem[Verner et al.(2003)]{verner03} Verner, E., Bruhweiler, F., Verner, D., Johansson, S., \& Gull, T. 2003, \apj, 592, L59
\bibitem[Verner et al.(1999)]{verner99} Verner, E. M., Verner, D. A., Korista, K. T., Ferguson, J. W., Hamann, F., \& Ferland, G. J. 1999, \apjs, 120, 101
\bibitem[V{\'{e}}ron-Cetty \& V{\'{e}}ron(2003)]{veron03} V{\'{e}}ron-Cetty, M.-P., \& V{\'{e}}ron, P. 2003, \aap, 412, 399
\bibitem[Vestergaard \& Peterson(2005)]{vestergaard05} Vestergaard, M., \& Peterson, B. M. 2005, \apj, 625, 688
\bibitem[Wills et al.(1985)]{WNW85} Wills, B. J., Netzer, H., \& Wills, D. 1985, \apj, 288, 94
\bibitem[Yoshii et al.(1998)]{yoshii98} Yoshii, Y., Tsujimoto, T., \& Kawara, K. 1998, \apj, 507, L113
\bibitem[Zheng et al.(1995)]{zheng95} Zheng, W., Kriss, G. A., Davidson, A. F., Lee, G., Code, A. D., Bjorkman, K. S., Smith, P. S., Weistrop, D., 
  Malkan, M. A., Baganoff, F. K., \& Peterson, B. M. 1995, \apj, 444, 632
\end{thebibliography}
\end{document}